\documentclass{aa}  
\usepackage{graphicx}
\usepackage{amsmath}	
\usepackage{amssymb}	
\usepackage{multicol}        
\usepackage{bm}		
\usepackage{pdflscape}	
\usepackage{xcolor}
\usepackage{float}
\usepackage{txfonts}
\newcommand{\ROSAT}{\emph{ROSAT}~}

\begin{document}

   \title{Runaways and shells around the CMa OB1 association}

   \author{B. Fernandes
          \inst{1,2}
          \and
          T. Montmerle
          \inst{1}
          \and 
          T. Santos-Silva
          \inst{1,2}
          \and
          J. Gregorio-Hetem
          \inst{2}
          }

   \institute{Institut d'Astrophysique de Paris, 75014, Paris, France
         \and
             Universidade de S\~ao Paulo, IAG, Departamento de Astronomia, S\~ao Paulo, 05508-090, Brazil\\
              \email{beatriz.fernandes@iag.usp.br}
             }

   \date{}

 
  \abstract
   {The origin of the arc-shaped Sh~2-296 nebula is still unclear. Mainly due to its morphology, the nebula has been suggested to be a 0.5 Myr-old supernova remnant (SNR) that could be inducing star formation in the CMa OB1
association. Therefore, this region can be an excellent laboratory for the investigation of the influence of massive stars on their surroundings.}
   {We aim to show, for the first time, that the nebula is part of a large, shell-like structure, which we have designated the ``CMa shell'', enclosing a bubble created by successive supernova (SN) explosions. We identified three runaway stars, associated with bow-shock structures, in the direction of the CMa shell and we investigate the possibility that they have originated in the center of the shell.}
   {By analyzing images of the CMa OB1 association at several wavelengths, we clearly see that the Sh~2-296 nebula is in fact part  of  a  large  structure, which can be approximated by a large (with a diameter of $\sim$ 60 pc) elliptical shell. Using the recent \textit{Gaia-DR}2 astrometric data, we trace back the path of the three runaway stars, in order to find their original position in the past, with relation to the CMa shell. We also revise the heating and  ionization of the Sh~2-296 nebula, by comparing the photon budget provided by the O stars in the region with results from radio observations.}
   {We find that the runaway stars have likely been ejected from a Trapezium-like progenitor cluster on three successive SN explosions having taken place $\sim$ 6, $\sim$ 2 and $\sim$ 1 Myr ago. We also show that the few late-type O stars in the region  cannot explain the ionization of the Sh~2-296 nebula and other mechanisms need to be at work.} 
   {We argue that, though we now have evidence for several SNe events in the CMa OB1 association, the SNe probably played a minor role in triggering star formation in these clouds. In contrast, the CMa OB1 association, as it is now, likely testifies  to the last stages of a star-forming region.}


   \maketitle
%

\section{Introduction} \label{sec:intro}

The stellar association CMa OB1 is composed of over 200 B stars and a few late-type O stars, located at a distance  $d \sim$ 1 kpc from the Sun \citep{Gregorio-Hetem2008}. Its most prominent feature is the extended arc-shaped Sh~2-296 (``Seagull'') nebula, part of the CMa R1 reflection nebulae association, known to be physically related to CMa OB1. It has long been suggested that the nebula is the remnant of an old supernova (SN) that could have triggered the formation of new stars in the region. Our current knowledge of the young stellar population in CMa OB1 hints to a complex star formation history, owing to the presence of young objects having originated in different star-formation events, with a low fraction of circumstellar disks, and a widespread evolved young stellar population of unknown origin. It is, therefore, a prime laboratory for the investigation of the influence of massive stars on their surroundings, in the past as well as at present. 

\cite{Herbst10} described CMa R1 as a partial emission ring with most association members embedded in an arc-shaped complex of dust clouds that seems to be  bound by ionized gas at its inner edge. These authors proposed a scenario in which the star-formation in the region would have been induced by a SN explosion about 0.5 Myr ago. They obtained this estimate by assuming that the arc is part of a SN remnant (SNR) in the late ``snowplow'' phase and expanding in a uniform medium \citep{Chevalier1974}. Later studies offered alternative origins for the nebula, such as the effects of strong stellar winds or an expanding old, ``fossil'' HII region \citep{Reynolds1978}. However, the main arguments in favor of a SNR still remain, including the large arc-shaped shell of emission observed in optical images, the near absence of obvious exciting stars, and also the presence of a runaway late O star (HD~54662), assumed to have been ejected from a close massive binary system after the explosion of its companion.

\cite{Reynolds1978} and \cite{Comer1998} also noted that the shell-like structure is in slow expansion (with velocities of 13 $\rm{km~s^{-1}}$ or $\simeq 20-30~\rm{km~s^{-1}}$, respectively). However, \cite{Comer1998} argue that even if a SN explosion took place in CMa OB1, there is strong evidence that star-formation was already occurring in pre-existing molecular clouds. \cite{Shevchenko1999} studied the stellar content of Sh~2-296 and found that most stars were much older (5 - 10 Myr) than the hypothetical time elapsed since the SN explosion, which therefore could not possibly have triggered the formation of these stars.

On the other hand, several sites of recent star formation are found in the nebula, such as the NGC~2327 and BRC~27 clusters, with ages of $\sim$ 1.5 Myr \citep{Soares2002}. The wide-field, but partial ($\simeq 2^{\circ} \times 1^{\circ}$) \ROSAT survey of the region by \cite{Gregorio-Hetem2009} identified a hundred stellar X-ray sources, particularly in two young star aggregates around the stars Z CMa and GU CMa, the latter previously unknown and located $\sim 30'$ away from the Z CMa cluster. Here, the IR characterization of the X-ray sources shows the coexistence of two groups of stars with different ages. Older (ages $>$ 10 Myr) and younger (ages $<$ 5 Myr) objects are found in both clusters and the authors proposed a possible mixing of populations in the ``inter-cluster'' region between these two groups of young stars.

As a follow-up of the \ROSAT observations, \cite{Santos-Silva2018} studied a sample of 387 X-ray sources detected with {\emph{XMM-Newton}. They found an even larger spread in ages and suggested that two episodes of star formation occurred in Sh~2-296. The first episode was slow and would have been completed over 10 Myr ago. A faster second episode is evidenced by the presence of a large number of objects younger than 5 Myr and disk-bearing T Tauri stars (TTSs) near the edge of the cloud \citep{Fernandes2015},
but these authors found a low fraction of disks among the young stars in Sh~2-296,  when compared to star-forming regions of similar ages. Therefore, the spatial segregation of disk-bearing stars at the edge of the nebula suggests that the nebula itself is somehow responsible for the early disappearance of disks, which might be the result of the mechanical (aerodynamic)  dissipation of disks due to past stellar winds or the passage of a shock wave from a SN explosion, or to some other SNR-related process.

Here, we will revisit these results in a broader context, in particular presenting new evidence showing that Sh~2-296 is an ionized region associated with multiple, nested SNRs and part of a closed shell-like structure having an age $\sim 6$  Myr, as indicated by the expanding proper motions of three runaway massive stars we find to be associated with the shell, and discuss several consequences.

The paper is organized as follows. In Sect. \ref{sec:runaways} we discuss the evidence for a large shell in CMa, and trace back the position of three runaway stars as clues to past supernova explosions spanning several million years. In Sect. \ref{sec:ionization}, the heating and ionization of the Sh~2-296 nebula are revisited, with the conclusion that, contrary to previous claims, its ionization can be explained only partially by the well-known existing O stars. 
In Sect. \ref{sec:discussion}, we put the results in a broader context, by discussing the plausibility of supernovae having triggered several episodes of star formation in the region. In this discussion, we stress the interest of the CMa bubble, which seems to be  rather unique when compared to other known galactic SNR in contact with molecular clouds. A general summary and final conclusions are presented in Sect. \ref{conclusions}.


\section{Runaway stars and evidence for a large shell} \label{sec:runaways}


\begin{figure*}
\begin{center}
\includegraphics[width=0.9\textwidth,angle=0]{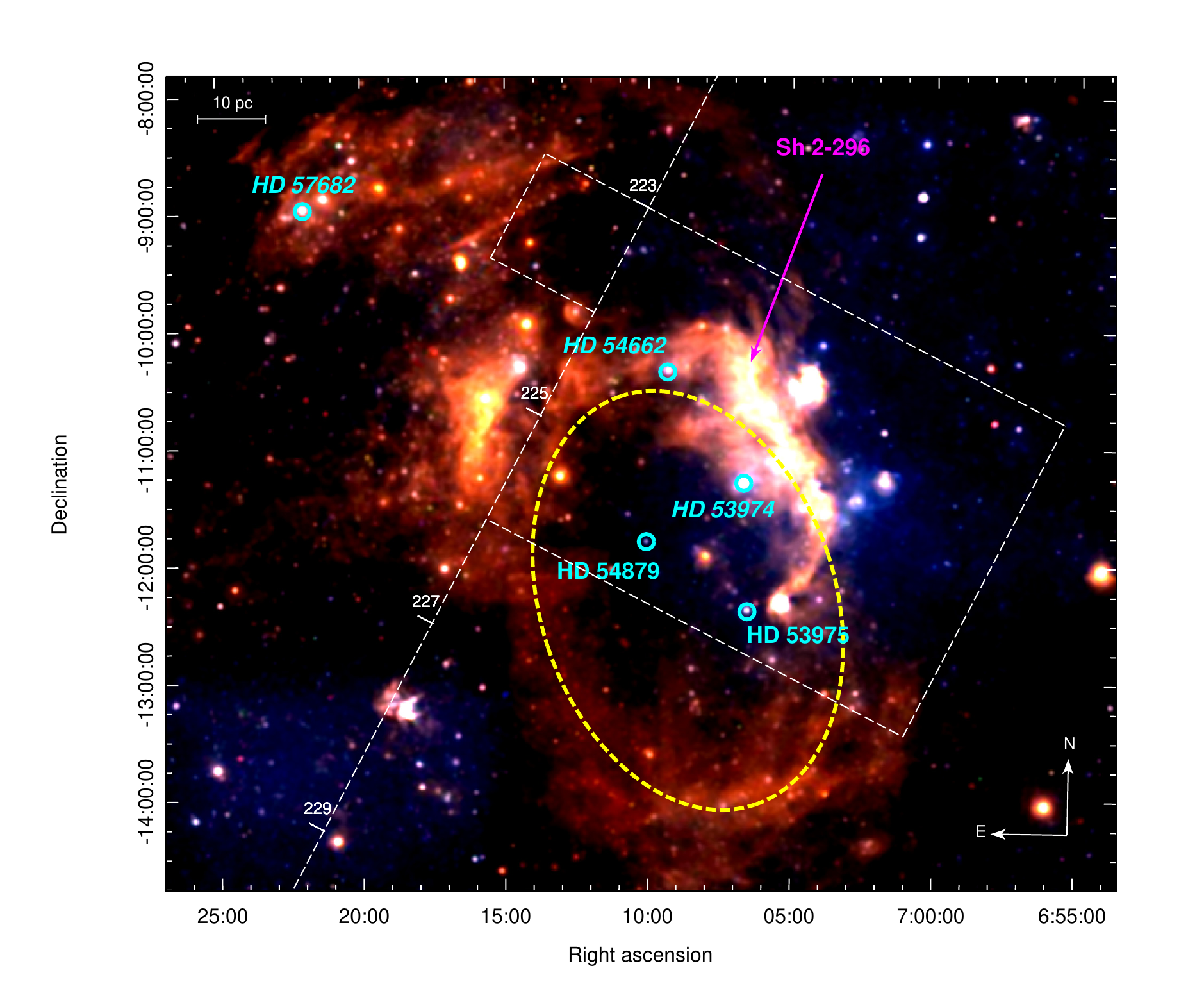}
\caption{RGB-color \emph{DSS} image of the region containing the CMa OB1 region and enlarged view of the Sh~2-296 nebula, extracted from the \emph{Progressive Sky (HiPS)} \citep{Fernique2015} tool from the \emph{Aladin Sky Atlas} \citep{Bonnarel2000}. The approximate position of three runaway stars (see Sec. 2.2), and of the other O star members of the association (HD~53975 and HD~54879),  are marked and labeled (the italics identify the runaways). The white thin lines trace the $^{13}$CO emission obtained with the \emph{Osaka} 1.85m mm-submm Telescope.  This emission is tightly correlated with the dust emission seen by the \emph{Planck} satellite (LFI instrument; see Fig. A1, which covers a larger area than this \emph{DSS} image), so we expect no CO emission outside of the \emph{Osaka} survey area (dashed box). We also show the coordinates on the galactic plane at $b = 0^\circ$. The ellipse in yellow marks the approximate shape of the CMa shell, centered at $\alpha$ = 07 08 36.824, $\delta$ = -12 17 41.26, with a semi-major axis of $\sim$ 32 pc at the adopted distance for the nebula of 1 kpc. The faint features visible at the bottom and at the top of the figure  (around HD~57682) are also visible in H$\alpha$ images and detected in the radio range (see Fig. \ref{arcs} and the discussion in Sect.3).}
\label{bubble}
\end{center}
\end{figure*}


In the wide-field ($9^{\circ} \times 9^{\circ}$) \emph{DSS} RGB image of CMa OB1 shown in Fig. \ref{bubble} it becomes apparent that the Sh~2-296 nebula is in fact part of a large structure  extending far to the East and South, which can be approximated by a closed, roughly elliptical shell  (the ``CMa shell'', hereafter). Indeed, such a shell may be associated with distinct features: in Fig.\ref{bubble}, the shape of the shell is delineated by the optical emission of the nebula and the molecular clouds distributed in its neighborhood, as recently observed in $^{13}$CO emission with the Osaka 1.85m mm-submm Telescope\footnote{Kindly provided by T. Onishi and K. Tokuda (private communication; see also \citep{Onishi2013} for a description of the telescope and the observational program.)}. Other structures are also visible in multiple wavelengths as we show in Appendix \ref{gallery} and the SE parts of the shell are detected in the 13 cm radio range (see Fig. \ref{contours}).

By visual inspection, we have fitted these features by an ellipse (yellow dashed ellipse shown in Fig. \ref{bubble} with a centroid at $RA(2000) = 07h~08m~36.82s$, $Dec(2000) = -12^{\circ} 17' 41.26''$, semi-major and semi-minor axes $a = 1.83^\circ$ and $b = 1.26^\circ$, which correspond respectively to $\sim 32$ pc and $\sim 22$ pc at the adopted distance for the nebula $d = 1$ kpc. Based on this approximation, the CMa-shell encloses a roughly ellipsoidal volume reaching $\sim$ 60 pc in size.


\begin{table*}
\centering
\caption{Data on the runaway stars associated with the CMa shell.}
{\small
\begin{tabular}{lll}\hline \hline 
Parameters & Value & Refs. \\ \hline
{\bf HD 53974} & B2I  &  Houk \& Swift (1990) \\
RA, Dec (J2000) & 07 06 40.77, -11 17 38.44 &  \\
l, b (deg) & 224.71, -01.79 &  \\
$\mu_{\alpha}$ cos $\delta $ (mas yr$^{-1}$) & -3.14 $^{\pm 0.72}$ & Van Leeuwen (2007)  \\
$\mu_{\delta}$ (mas yr$^{-1}$) & 3.32 $^{\pm 0.55}$ & Van Leeuwen (2007) \\
Parallax (mas) & 1.07 $^{\pm 0.61}$ & Van Leeuwen (2007) \\
Radial velocity (km s$^{-1}$) & 31.0 $^{\pm 4.2}$ & Gontcharov (2006) \\ \hline
{\bf HD 54662} & O7V & Sota et al. (2014) \\
RA, Dec (J2000) & 07 09 20.25, -10 20 47.63 &  \\
l, b (deg) & 224.17, -0.78 &  \\
$\mu_{\alpha}$ cos $\delta $ (mas yr$^{-1}$) & -2.055 $^{\pm 0.142}$ & Gaia Collaboration et al. (2018) \\
$\mu_{\delta}$ (mas yr$^{-1}$) & 2.645 $^{\pm 0.169}$ & Gaia Collaboration et al. (2018) \\
Parallax (mas) & 0.85 $^{\pm 0.51}$ & Gaia Collaboration et al. (2018) \\
Radial velocity (km s$^{-1}$) & 57.9 $^{\pm 2.5}$ & Gontcharov (2006)  \\ \hline
{\bf HD 57682} & O9.2IV & Sota et al. (2014) \\
RA, Dec (J2000) & 07 22 02.05, -08 58 45.76 &  \\
l, b (deg) & 224.41, +02.63 &  \\
$\mu_{\alpha} $ cos $\delta $ (mas yr$^{-1}$) & 9.772 $^{\pm 0.089}$ & Gaia Collaboration et al. (2018) \\
$\mu_{\delta}$ (mas yr$^{-1}$) & 12.821 $^{\pm 0.074}$ & Gaia Collaboration et al. (2018) \\
Parallax (mas) & 0.81 $^{\pm 0.59}$ & Gaia Collaboration et al. (2018) \\
Radial velocity (km s$^{-1}$) & 24.1 $^{\pm 1.2}$ & Gontcharov (2006)  \\ \hline
\label{dados}
\end{tabular}
}
\end{table*}


\begin{figure}
\includegraphics[width=0.48\textwidth,angle=0]{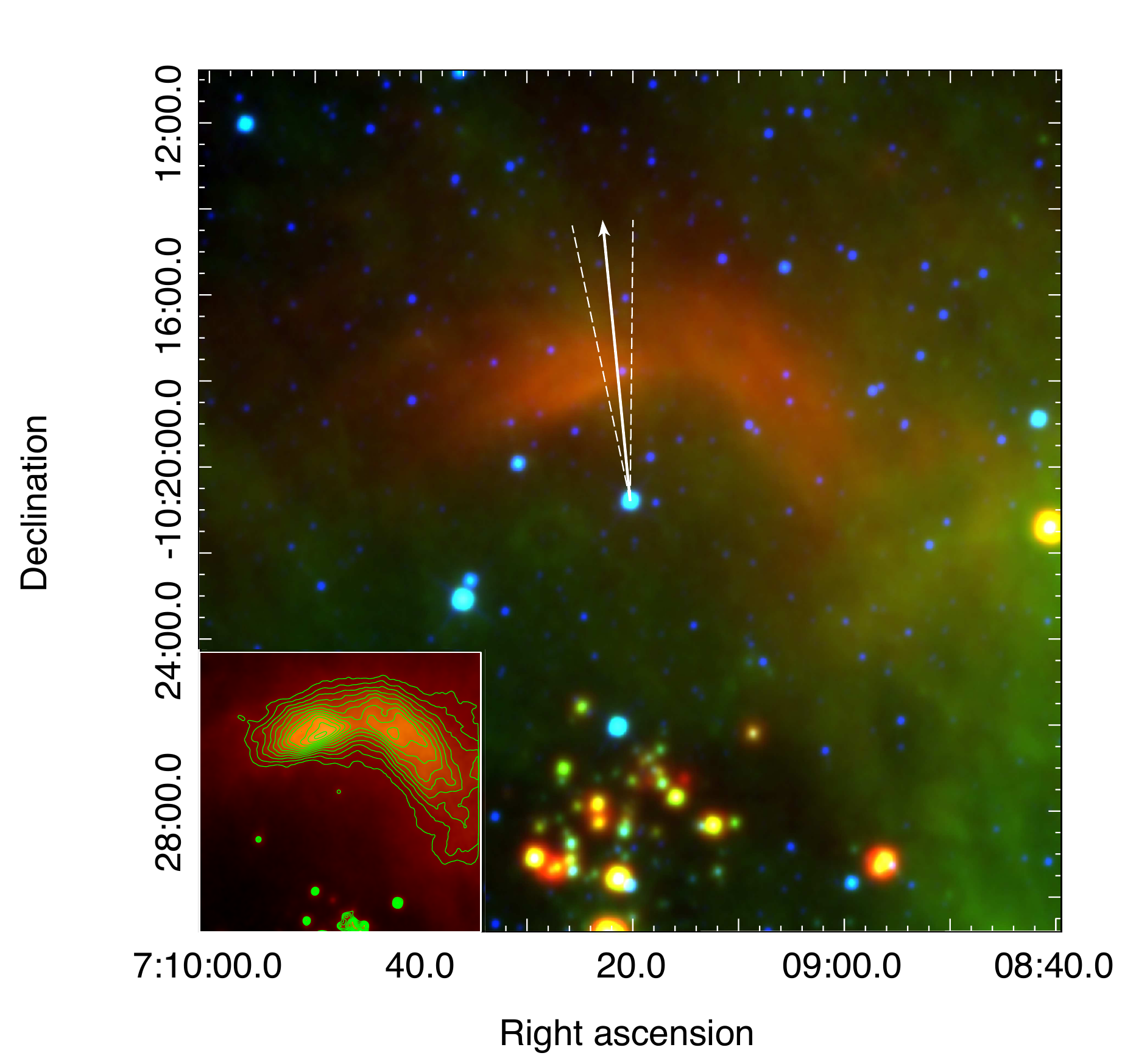}
\caption{\emph{WISE} composite color image of the bow-shock associated with HD 54662. Red: band 4, 22.2 $\mu$m. Green: band 3, 12.1 $\mu$m. Blue: band 1, 3.4 $\mu$m (North is up, East to the left). The white vector (not to scale) represents the direction of the proper motion   \citep{GaiaDR22018}, corrected for galactic rotation and basic solar motion, and the 1$\sigma$ errors are represented by dashed lines. A \emph{WISE} 22 $\mu$m image of the bow-shock is inset at the bottom left of the figure, overlaid with linearly-scaled contours of 22 $\mu$m emission. 
Note that the velocity vector is oriented towards the densest region of the bow-shock, not towards its apparent apex.
}
\vspace{-0.5cm}
\label{bowhd54662}
\end{figure}



\begin{figure}
\includegraphics[width=0.48
\textwidth,angle=0]{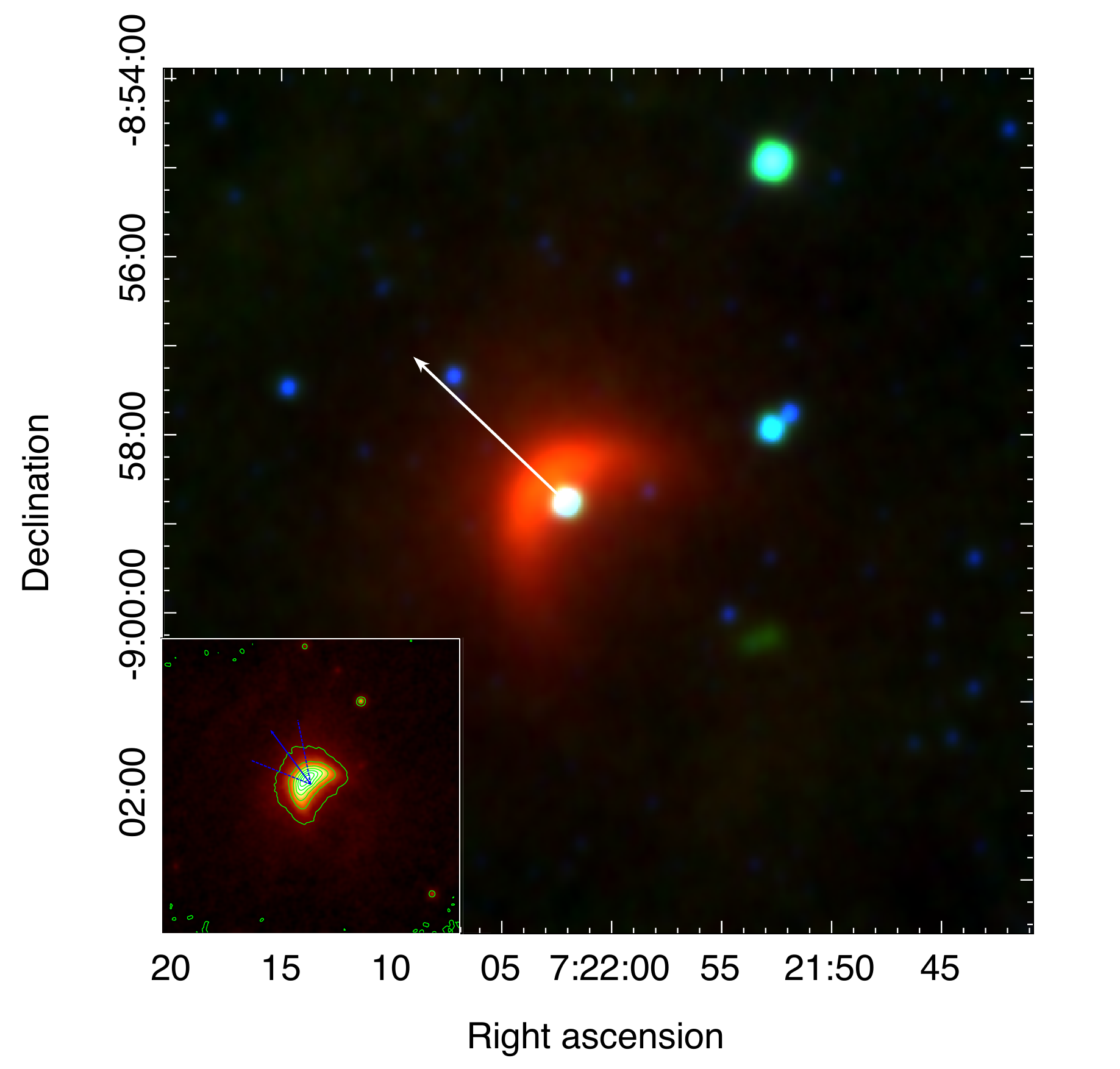}
\caption{Same as Figure \ref{bowhd54662}, but for HD 57682. In the insert, the blue vectors indicates the preferential orientation of the bow-shock and 1$\sigma$ errors, as outlined by the contours. The vectors lengths are not to scale (North is up, East to the left).}
\label{bowhd57682}
\end{figure}


Three stars associated with bow-shock structures are found in the direction of the CMa shell. Bow-shocks can often be the result of the interaction of the winds of massive runaway stars with their local interstellar medium. Two scenarios can explain the high space velocities of runaway stars: the dynamical ejection of the star from a dense cluster \citep{Poveda1967} or the supernova explosion of a close, more massive, binary companion \citep{Blaauw1961}. 

These stars all have parallaxes that agree with the adopted distance of CMa OB1 ($d \sim 1$ kpc) within 1$\sigma$, they appear to be moving away from its center, and are now located inside or even beyond the shell. In the absence of such a dense massive-star cluster inside the CMa shell, the most-likely scenario to explain the presence of runaways is the SN binary disruption.

The first runaway has been known for some time: it is the fast-moving HD~54662, an O7V star identified by \cite{Herbst10} as a confirmed member of the association (see Fig. \ref{bubble}). They noted that the star has a radial velocity which differs from the mean value for the association by $\sim$ 30 km~s$^{-1}$ and proposed that its peculiar velocity could be the result of an ejection resulting from a SN explosion in a binary system at approximately 0.5 Myr ago. As noted by \cite{Fierro} this is a peculiar object that has been treated as a binary by some authors \citep{Fullerton, Sana, Mossoux}, and as a single star by others \citep{Markova,Krticka}.  \cite{Fierro} find no binarity evidence in the spectra of the star. Here, we will also treat it at first as a single runaway star, but discuss the implication of its possible binarity on the ionization of the Sh~2-296 nebula in Sect. \ref{sec:ionization}.

A second high-velocity star is also found further North of the nebula  (see Fig. \ref{bubble}), HD~57682. This is a fast-proper motion O9.2IV star \citep{Sota2014} that has been previously suggested by \cite{Comer1998} to be a runaway star possibly related to a violent formation of the CMa OB1 complex. 

These two runaway stars have created bow-shocks, seen in the mid-IR range by the $WISE$ survey  \citep{Wright2010} observations, as a result of their supersonic motion in the ambient Interstellar Medium (ISM), with a velocity oriented along the vertex of the bow shape (see Figs. \ref{bowhd54662} and \ref{bowhd57682}).

We found a third star associated with a bow-shaped structure located near the inner edge of the CMa shell, HD~53974. The bow-shock was detected in $WISE$ catalogue images \citep[see][]{Peri2015}, and is also clearly seen in H$\alpha$, in front of the star trajectory (see Fig. \ref{figbows}), which moves towards Sh~2-296. Both HD~53974 and HD~54662 are listed in the bow-shock catalogue of \cite{Kobulnicky2016} as isolated bow-shocks, which the authors 
associate with genuine runaway stars. The Herbig Ae/Be star HD~53367 (exciting the IC~2177 HII region, part of CMa OB1) also appears in this catalogue, but it is classified by the authors as an ``in-situ'' bow-shock and not associated with a runaway.

In Table \ref{dados}, we list the relevant data for HD~53974, HD~54662 and HD~57682 obtained from the literature. We adopted the recent \emph{Gaia DR2} proper motions \citep{GaiaDR22018} for the stars HD~54662 and HD~57682, while for HD~53974, only the less accurate proper motions from \emph{Hipparcos} \citep{vanLeeuwen2007} are available.


\begin{figure*}
\centering
\includegraphics[width=0.48\textwidth,angle=0]{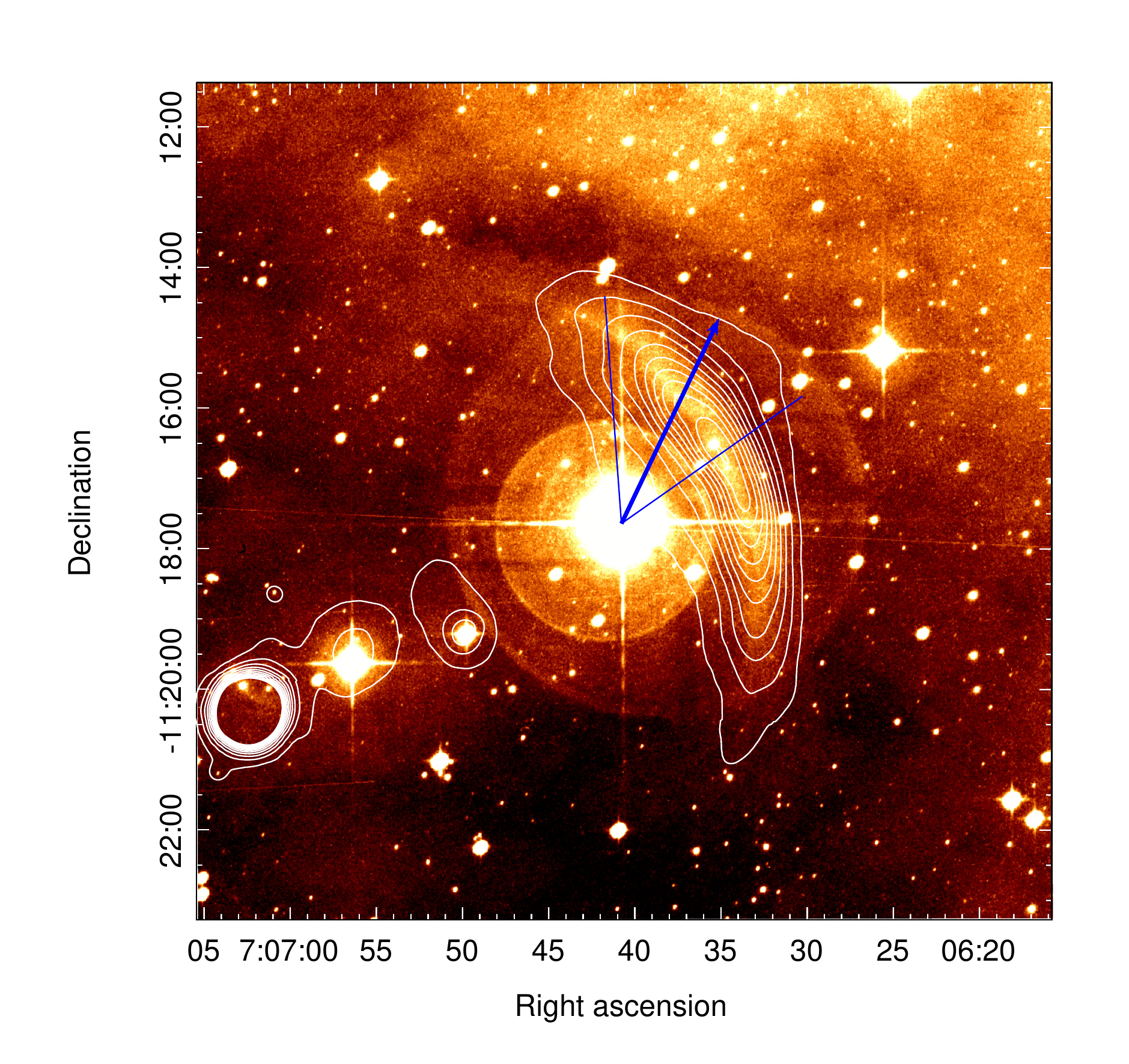}
\includegraphics[width=0.48\textwidth,angle=0]{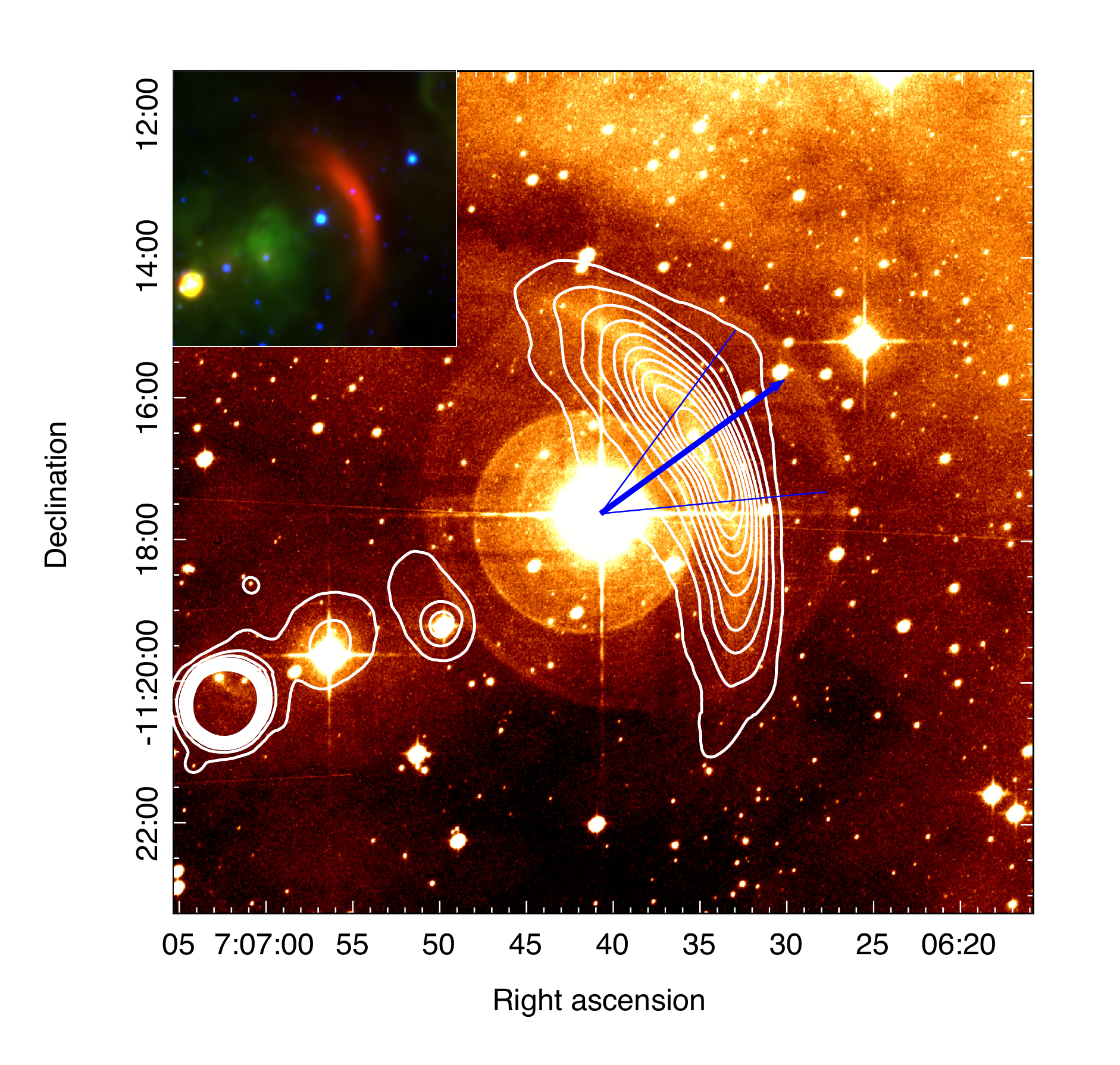}
\caption{Images of the bow-shock associated with HD 53974 from the AAO/UKST SuperCOSMOS H$\alpha$ survey \citep{Hambly2001} overlaid with linearly-scaled contours of the emission at 22.2 $\mu$m that outline the shape of the bow-shock. {\it Left:} The blue vector represents the direction defined by the \emph{Hipparcos} proper motion corrected for galactic rotation and basic solar motion, within 1$\sigma$ errors. {\it Right:} A \emph{WISE} color-composite image of the bow-shock is inset at the top left corner of the figure. The blue vector indicates the preferential orientation (apex) of the bow-shock, which we define by drawing a straight line from the location of the star to the point of highest 22 $\mu$m emission. The 1$\sigma$ uncertainty (thin blue lines) is outlined by the highest level contour from the 22.2 $\mu$m emission. The vectors lengths are not to scale (North is up, East to the left).}
\label{figbows}%
\end{figure*}


To obtain the peculiar velocities of the stars, which indicate their motion with respect to their ambient ISM, we corrected the observed heliocentric velocities for solar motion and galactic rotation \citep[see e.g.][]{scheffler1988,moffat1998,Comeron2007}. 
For the calculations we assume a flat rotation curve with the Sun’s galactocentric radius  $R_0$ = 8 kpc, and velocity $V_0$ = 230 km s$^{-1}$, and the basic solar motion ($U_{\odot}$, $V_{\odot}$ , $W_{\odot}$) = (11.1, 12.24, 7.25) km~s$^{-1}$, as given by \cite{Schonrich2010}. 
We find the peculiar proper motions and velocities listed in Table \ref{pmres}. Note that the orientation (angle $\theta$) is given with respect to the galactic North (counted counterclockwise).  In the equatorial coordinates system, as shown in Figs. \ref{bowhd54662}, \ref{bowhd57682}, and \ref{figbows}, the vector orientation is  $\theta  - 23.5^o$, counted from North to East. Detailed calculations are presented in Appendix B. For the error estimate presented in Table \ref{pmres}, only the errors of the proper motion and the radial velocity measurements were considered.  We have checked that large errors on proper motion affect more the results than distance uncertainties. By adopting the error bars of parallax measured by {\it GAIA-DR2} for HD~54662 and HD~57682 the deviation on $\theta$ would be 23$^o$,  and 2.2$^o$, respectively.

We see that in the particular case of HD~53974, for which we only have \emph{Hipparcos} data, the direction of the derived peculiar proper motion is not well aligned with the observed vertex of its corresponding bow-shock, as illustrated in Fig. \ref{figbows}. Taking into account that the angular uncertainties on the proper motions alone can be very large, as is the case for instance for the fast runaway BD+43$^\circ$3654 studied by \cite{Comeron2007} in the Cyg~OB2 association, we choose to make use of the extra information provided by the bow-shock morphology to better constrain the direction of motion. 
Following the methodology described by \cite{Sexton2015}, we define the vector orientation of the motion for HD~53974 by drawing a straight line from the location of the star to the point of brightest 22 $\mu$m emission, taken to be the apex of the bow-shock (Note that the same situation holds for the other two runaways). Adopting the same reference used for the proper motion vector,  the angle $\theta_{{\rm bow-shock}}$ (given in Table \ref{pmres}) is the orientation of the bow-shock axis, with respect to the galactic North (counted counterclockwise). 

Tracing back the positions of the three runaways, as shown on Fig. \ref{traceback},  we find that all three stars would have been nearly at the same location, with their trajectories mutually crossing (near the center of the shell, close to the current positions of HD~54361\footnote{HD~54361 (W CMa) is a massive carbon AGB star for which the parallax (p" = 1.805 mas) and proper motion ($\mu_{\alpha}$ = -6.5 mas/yr; $\mu_{\delta}$ = +2.3 mas/yr) have been recently reported by \emph{GAIA DR2}. These data, combined with the estimate radial velocity ($\sim$ 19-23 km/s) for HD~54361, are evidence that its position at the center of the shell may be just a coincidence.} and of the LBN1038 nebula) $\sim$ 1 Myr ago for HD~57682, $\sim$ 2 Myr ago for HD~53974 (within the large \emph{Hipparcos} velocity vector errors), and $\sim$ 6 Myr ago for HD~54662.  In turn, this finding strongly suggests that the three runaways originally belonged to a compact cluster of massive O stars inside the shell.

\begin{table*}
\centering
\caption{Peculiar proper motions of runaway stars in CMa OB1. The angle $\theta$ gives the orientation (counterclockwise with respect to galactic N) of the velocity vector. The same as for $\theta_{{\rm bow-shock}}$, which is based on the arc-shape of the 22 $\mu$m emission.}
\begin{tabular}{lcccccc}\hline \hline 
Star	&	${\mu_l}_{\rm{pec}} \cos b$	&	${\mu_b}_{{\rm pec}}$ & $| \mu |_{{\rm pec}}$ & $\theta$ & $\theta_{{\rm bow-shock}}$  & $v_{\star}$ \\
   &  (mas yr$^{-1}$) & 	(mas yr$^{-1}$) & (mas yr$^{-1}$)  & (deg) & (deg) & (km s$^{-1}$)\\ \hline
HD 53974	&	-1.94 $^{\pm 0.16}$ 	&	0.06 $^{\pm 0.68}$ & 1.94$^{\pm 0.37}$	& 1.87 $^{\pm 29.70}$& $-26.04_{-30.70}^{+17.20}$  & 12.28 $^{\pm 4.84}$\\
HD 54662	&	-0.95 $^{\pm 0.11}$	&	0.62 $^{\pm 0.20}$ & 1.14$^{\pm 0.18}$ & 33.16 $^{\pm 6.28}$ & -- & 36.30 $^{\pm 2.63}$ \\
HD 57682	&	-4.37 $^{\pm 0.03}$	&	16.14 $^{\pm 0.11}$	& 16.72$^{\pm 0.12}$ & 74.84 $^{\pm 0.02}$ & 64.65 $^{\pm 24.13}$ & 134.02$^{\pm 5.44}$ \\ \hline
\label{pmres}
\end{tabular}
\end{table*}

Going one step further, we propose a scenario in which three successive SN explosions, having taken place $\sim 1$ to $\sim 4$ Myr apart, would have disrupted three distinct O-star binary systems belonging to
an Orion-like, hierarchical cluster. For comparison, it is well known that three runaway stars ($\mu$ Col, AE Aur and 53 Ari), which have similar spectral types (resp. O9.5V, O9.5V, and B1.5V; compare with CMa, Table \ref{dados}), and possibly others, likely have a common origin in the Trapezium cluster \citep{Poveda2005}. Also, it is worth adding that Barnard's Loop, which is seen as a faint H$\alpha$ half-circular feature, centered on the Orion Trapezium, has also been interpreted as  nested remnants of several SN explosions (See Sect. 4.2). So the scenario we suggest is not unique and is in fact quite plausible.

\ \ 


\begin{figure*}
\begin{center}
\includegraphics[height=0.9\textwidth,angle=0]{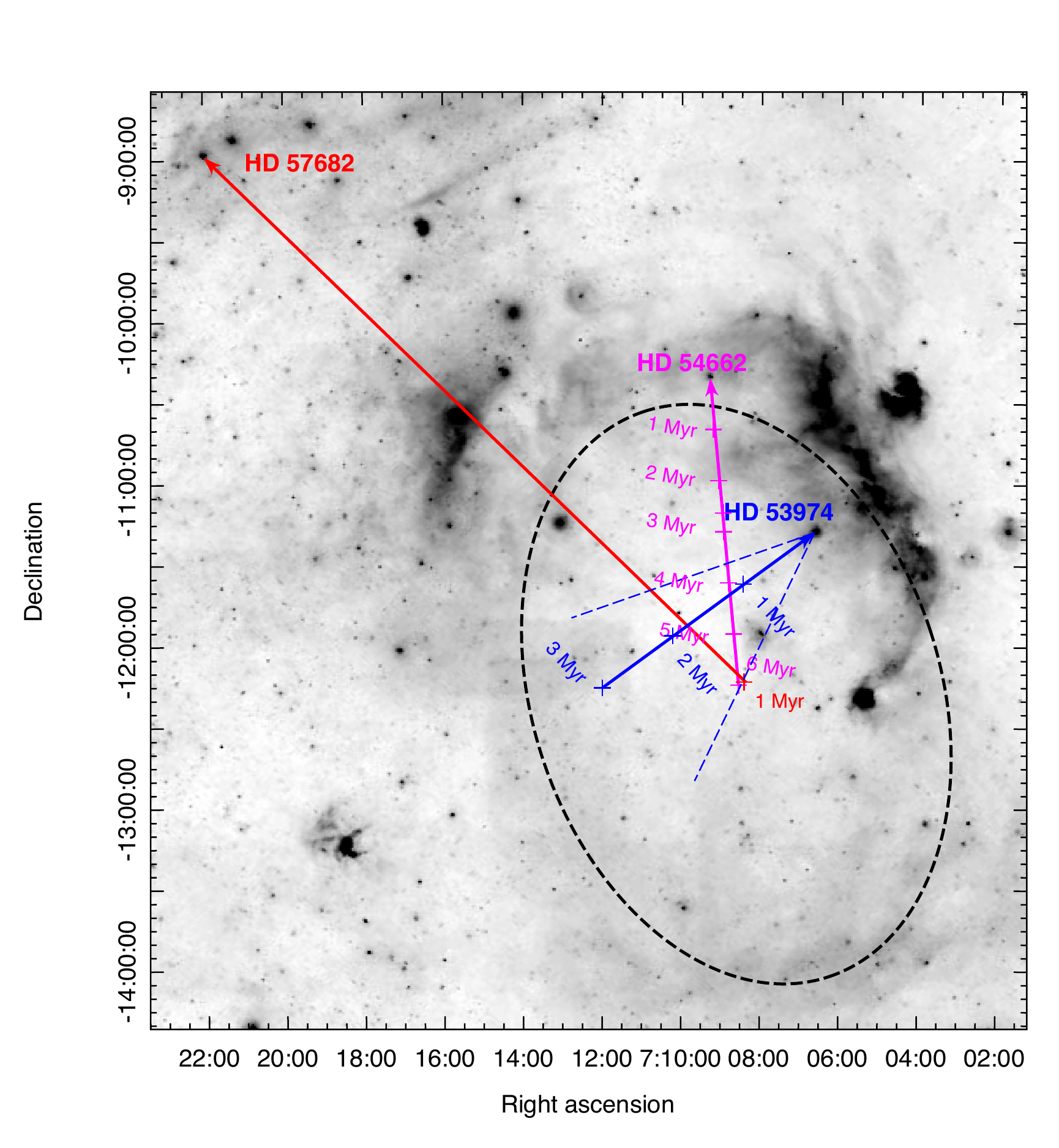}
\caption{\emph{DSS2}-R image of the region containing CMa R1 (North is up, East to the left). The ellipse shows the approximate extent of the shell and the arrows indicate the past positions of the runaway stars going back a few Myr (as labeled in the image), according to the derived proper motions of HD~57682 and HD~54662, and the bow-shock orientation of HD~53974. The dashed blue lines, represent the 1$\sigma$ uncertainty on the axis of the bow-shock associated with HD~53974 as discussed in the caption of Fig. \ref{figbows}. The stars HD~54662, HD~53974 and HD~57682 would have been at a coinciding position, located close to the center of the CMa shell, at respectively $\sim 6$ Myr, $\sim 2$ Myr and $\sim 1$ Myr ago.}
\label{traceback}
\end{center}
\end{figure*}


\section{The source of ionization of the Seagull Nebula} \label{sec:ionization}

The ionization of the Sh~2-296 nebula and its related components (smaller nebulae, etc.) is clearly noticed at many wavelengths. The most detailed information can be extracted from H$\alpha$ images, and in the radio cm range. This information can then be crossed with information about the dense, cold material content, such as maps obtained at 21 cm (HI), infrared (dust) and CO (tracing H$_2$).

\subsection{Structure and morphology of the ionized gas} \label{sec:structure}

Figure \ref{halphazoom} shows that the H$\alpha$ emission is brightest along an approximately linear ridge running NE-SW. A closer look shows that the ridge is structured with locally brighter emission regions, but otherwise has a fairly uniform brightness, until it bends to the S to its tip, the small Sh~2-297 nebula. To the N, Fig. \ref{contours}a shows increasingly extended, bright thin filaments, up to $\sim 10$ pc long, that appear to be running approximately perpendicular to the ridge, and make up a large-scale, diffuse H$\alpha$ emission component (see also Fig.\ref{halphazoom} for more details).
The bright nebula IC2177 (Sh 2-292), also called “the Head of the Seagull”, is located to the NW of the ridge.

 In Fig. \ref{contours}a we use the same DSS2-R background image of Fig. \ref{traceback}, which corresponds to the H$\alpha$ emission. When the $^{13}$CO contours (displayed in Fig. \ref{contours}b) are superimposed on this optical background, some striking features become apparent. $(i)$ A ``chain'' of molecular clouds closely follows the ridge, and is dominated by a large, elongated cloud which deviates to the W from the ridge towards and including the IC2177 nebula: we will refer to it as the ``West Cloud'', which has an estimated mass of $M_W \sim 1.6 \times 10^4 M_{\odot}$ \citep{Kim2004}. These molecular clouds are associated, with active, ongoing star formation and young stars. $(ii)$ To the E of this cloud is found a separate, also extended but weaker CO emission  that we will refer to it as the ``East Cloud'', which mass is $M_E \sim 1.2 \times 10^4 M_{\odot}$ \citep{Kim2004}. In contrast to the West Cloud, this cloud is not associated with strong H$\alpha$ emission, but it fills extremely well the ``hole'' near NGC~2343 (visible on the zoom of the nebula, Fig. \ref{halphazoom}), and even the structure with low surface brightness that goes to the E. 

The West and East Clouds are also traced by the visual extinction (A$_V$) map (Fig. \ref{contours}c), produced by \cite{Dobashi2011} based on data from \emph{2MASS Point Source Catalog} (2MASS PSC). We can see that the dust closely follows the $^{13}$CO distribution, therefore we can infer from the $A_V$ contours that the ``chain'' of molecular clouds extends to the south-east of the region covered by the $^{13}$CO survey, following the approximate shape of the CMa shell. In addition, the dark filaments (seen in Fig. \ref{halphazoom}) also fit extremely well inside the A$_V$ and $^{13}$CO contours that we see in Fig. \ref{contours}. This indicates that the East Cloud must be situated {\it in front} of the diffuse bright filamentary component. Note that, in spite of the spatial correspondence, there is no link between this cloud and NGC~2343, which has an estimated age of $\sim$ 60-100 Myr \citep{Kharchenko2005a}.

We shall see below that this cloud plays a crucial role in the interpretation of the ionization structure of the nebula.


\begin{figure}
\centering
\includegraphics[width=0.48\textwidth,angle=0]{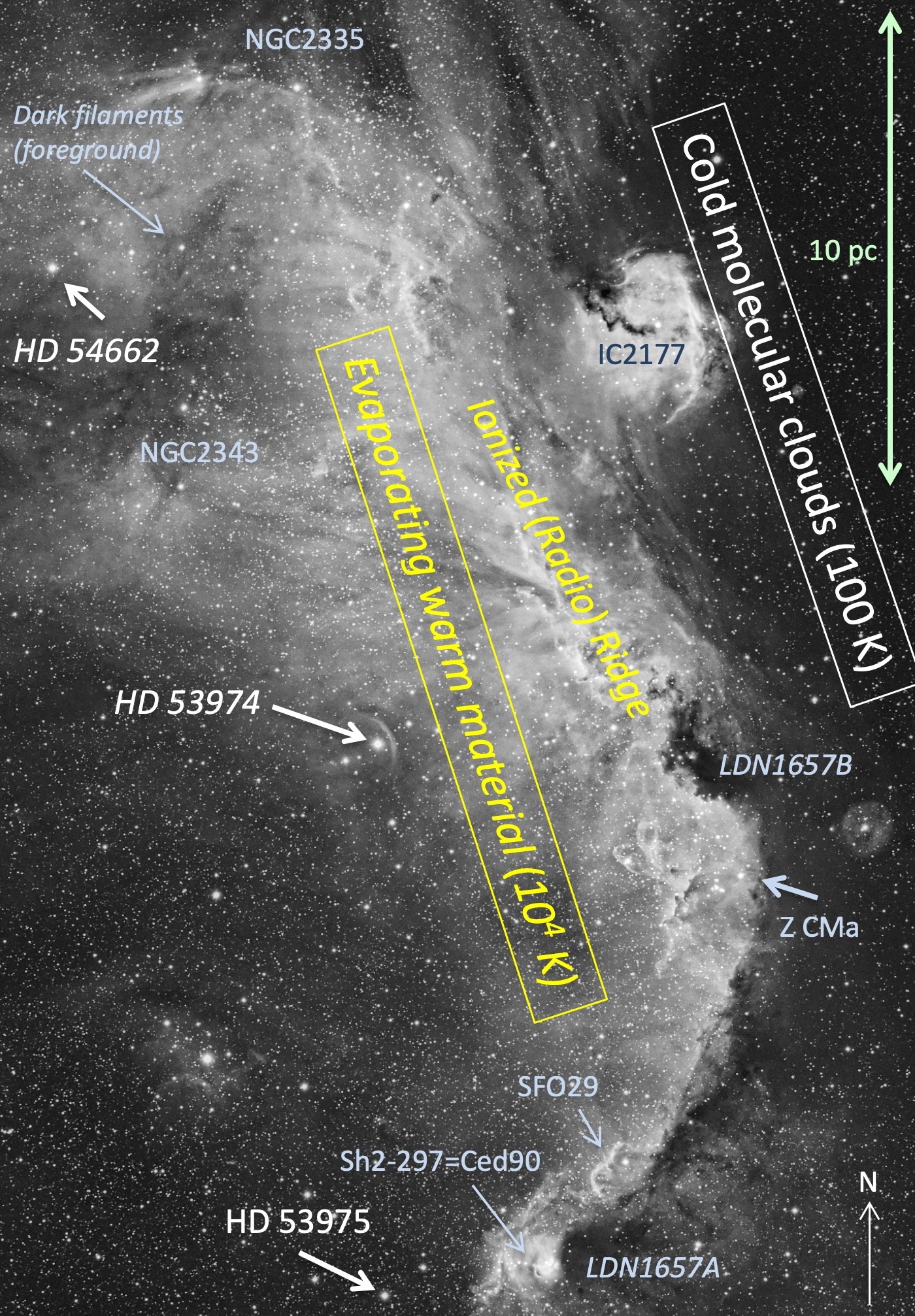}
\caption{Deep, close-up view of the Sh~2-296 nebula in H$\alpha$ with significant labels added (image courtesy of S. Guisard (www.sguisard.astrosurf.com) \& T. Demange (www.capnature.com )). This view vividly illustrates ionized evaporation features flowing away from the adjacent molecular clouds: long (up to $\sim 10$ pc) filaments, smaller-scale ``eggs'' and ``pillars'' (as, for example, near the Herbig Be star Z~CMa, or the SFO~29 filament) and bright, dense corrugations. Interacting with this flow, the runaway O star HD~53974 has created a conspicuous bow shock (see text, Sect. 2.2, and Fig.\ref{figbows}). Given the evaporation features it is plausible that the bubble edge is actually located {\it behind} the Lynds dark nebulae LDN1657A and LDN1657B.}
\label{halphazoom}
\end{figure}


As shown in Fig. \ref{contours}b, the region in between the two CO clouds is also associated with bright H$\alpha$ emission, but the gas may be too warm there to be visible in $^{13}$CO. It could be warmer, less dense HI gas, but the 21-cm data are too coarse to allow a detailed study of the spatial correspondence. Nevertheless, the bright 21-cm emission shown in Appendix A does overlap the ``intercloud'' region, so material is indeed present in spite of the absence of $^{13}$CO emission.

Another way of studying ionized gas is to look for radio emission, i.e., continuum and/or recombination lines. 
\cite{Nakano1984} have used the Nobeyama 45-m telescope for continuum observations of the nebula at 10.2 GHz (resolution $2.7'$). In Fig. \ref{contours}d, we have superimposed the radio contours on the \emph{DSS}-red image. Given the good spatial resolution, the 10 GHz emission is seen to follow very closely the H$\alpha$ ridge brightness variations, including the absence of a large-scale gradient along the ridge, and part of the diffuse emission at lower levels. In other words, this radio emission is an excellent probe of the ionization front of the nebula and its close vicinity. Other features, such as the IC~2177 and Sh~2-297 nebulae, also have an excellent correspondence in the radio. Quantitatively, \cite{Nakano1984} derive an average electron density for the HII gas $n_{e,i} \simeq 10~{\rm cm}^{-3}$, reaching $n_{e,i} \sim 50~{\rm cm}^{-3}$ in bright, localized sources. 

On the other hand, \cite{Gaylard1984} have used the Hartebeesthoek 26-m telescope at 13 cm (or 2.3 GHz; HPBW = $20'$), to survey a  much larger area ($\sim 8^{\circ} \times 10^{\circ}$) around the nebula. They also observed the H$142\alpha$ recombination line, to extract velocity information. The coarser resolution however does not allow a precise matching of morphological details. The authors find an average electron density $n_{e,i} \simeq 10-20~{\rm cm}^{-3}$, consistent with the result found by \cite{Nakano1984}. Coupling their results with other radio results, they were able to find the spectral index, indicating thermal bresstrahlung emission with $T_{e,i} = 6900 \pm 1300$ K. The H$142\alpha$ observations yielded a turbulent velocity $v_t \sim 10 \pm 8~{\rm km~s}^{-1}$. This velocity is comparable to the sound velocity $c_s = (2kT_{e,i}/m_p)^{1/2} \simeq  10~{\rm km~s}^{-1}$, which means that there is no evidence for an expansion of the ionized gas.

\subsection{Stellar ionization vs. Lyman continuum emission} \label{sec:lymancontinuum}

The standard method to explain the ionization of nebulae such as Sh~2-296 is to establish a census of the ionizing O and B stars present in their vicinity, more or less facing the ionization front. The surveys of \cite{Claria1974a} and \cite{Shevchenko1999} are complete for stars brighter than V = 9.5 mag within the CMa OB1 association, indicating that there are no other O-type stars than the four that we list in Table \ref{lymanflux}. The corresponding Lyman continuum flux $Q_0$ (number of ionizing photons per second) depends on the spectra of the hot stars, and has been estimated by various authors \citep[e.g., ][]{Panagia1973, Schaerer1997, Martins2005}. Table \ref{lymanflux} gives a census of the
relevant O and B stars, with revised spectral types and recent estimates of the O-star Lyman continuum fluxes,  in particular taking into account
mass loss and line blanketing \citep{Martins2005}. Note that \cite{Martins2005} do not give $Q_0$ for B stars, but the previous study by \cite{Schaerer1997}  shows that they contribute in a negligible way to the total Lyman continuum flux ($< 10^{48}$ s$^{-1}$). We note also that two nebulae have B0V stars as their own ionization sources \citep[Sh~2-297,][and IC~2177]{Mallick2012}, so that we can remove them from the total observed Lyman continuum flux.

Considering only the three late O stars close to Sh~2-296, from Table \ref{lymanflux}, we see that  together contribute with $Q_0 = 1.05 \times 10^{49}$ s$^{-1}$. We can compare this value with the estimate from the radio observations of \cite{Nakano1984}, who derive a Lyman continuum flux $N_{Lyc} \simeq 1.3 \times 10^{49}~{\rm s}^{-1}$, integrated over the ridge (see Figs.\ref{contours}a and \ref{contours}d), and $N_{Lyc} \simeq 1.8 \times 10^{49}~{\rm s}^{-1}$, in total, integrated over distinct sources like IC~2177 and Sh~2-297. \cite{Gaylard1984} also give a consistent estimate of the total Lyman continuum flux $N_{Lyc} = 1.6~(+1.4/-0.8) \times 10^{49}~{\rm s}^{-1}$. However, if HD~54662 is really a (spectroscopic)  binary \citep{Sota2014}, the total ionizing flux rises to $Q_0 = 1.2 - 1.7 \times 10^{49}$ s$^{-1}$. Given this uncertainly, strictly speaking the estimated range of $Q_0$ is indeed compatible with, albeit lower than, the values of $N_{Lyc}$ deduced from the radio observations of both \cite{Nakano1984}  and \cite{Gaylard1984}.


\begin{table}
\caption{Computed Lyman continuum flux of the hot  stars of the CMa OB1 association. (Runaways are indicated in italics.)} 
\begin{center}
\begin{tabular}{lcc} \hline \hline
Star  & Sp. Type  & log$Q_0^{(1)}$ \\
 & (SIMBAD) &  (${\rm s}^{-1}$) \\
\hline
{\it HD~54662}$^{(2)}$   &  O6.5 V $^{\textrm{(a)}}$  & 48.88 \\
HD~53975   &    O7.5V z$^{\textrm{(c)}}$   & 48.61 \\
{\it HD~57682$^{(3)}$}  &   O9.2IV$^{\textrm{(b)}}$   &  47.97  \\
HD~54879   &    O9.7V$^{\textrm{(c)}}$     & 47.88 \\
{\it HD~53974}              &    B2I a/ab$^{\textrm{(d)}}$   & -- \\
HD~53755   &   B1II/III$^{\textrm{(d)}}$      & -- \\
HD~53456   &    B1Ib/II$^{\textrm{(d)}}$    & -- \\
HD~53367$^{(4)}$    &    B0IV/Ve $^{\textrm{(e)}}$   & -- \\
HD~53623$^{(5)}$ & B0V $^{\textrm{(f)}}$ & -- \\
\hline
\hline
\end{tabular}
\end{center}
\label{lymanflux}

\small

$(1)$ Values from \cite{Martins2005}, Table 4.
$(2)$  This runaway could be a spectroscopic binary, both with spectral types O6.5V, according to Mossoux et al (2018).
$(3)$  This runaway is $\sim 5^\circ$ away from the nebula (see Fig.\ref{bubble}).
$(4)$ Ionizes the IC~2177 nebula.
$(5)$ Ionizes the Sh~2-297 nebula.

\noindent
References for spectral types: (a) Mossoux et al. (2018) (b) \cite{Sota2014}; (c) \cite{Sota2011}; (d) \cite{Houk1999}; (e) \cite{Tjin2001}; (f) \cite{Houk1988}.
 
\vspace{0.2cm}
\normalsize
\label{lymantable}
\end{table}


Nevertheless, the largest contributor of Lyman continuum photons,  the runaway HD~54662, deserves further discussion. Indeed, the line of sight to this star happens to be grazing the East Cloud, yet no ionization is found in or around the cloud. Since both the East and West clouds have the same radial velocities, there is little doubt that they are part of the same complex, associated with CMa OB1, and we would expect both clouds to display comparable ionization features (H${\alpha}$ emission, radio emission, etc.). But in reality, we see that while the West Cloud, and its HI extension to the N, has a sharp ionization front, the East Cloud (and more specifically its dense dust filaments)  simply absorbs the associated, background diffuse HI emission coming from the West Cloud.  In addition, we have already mentioned that there is no evidence for a gradient of ionization along the ridge -- in other words, no $1/r^2$ distance effect from HD~54662 (or its proposed companion) on the ionization of the West cloud, which would be expected if it were the ionization source. As it turns out, the \emph{Gaia DR2} proper motions (Table \ref{pmres}) indicate that the star would be now $\sim 200$ pc farther away from its original location 6 Myr ago, meaning that the star would be currently ``isolated'' in a distant, low-density region of the ISM, too far to significantly ionize the East Cloud, and by the same token, too far from the West Cloud (and its associated, inter-cloud HI component). Therefore, HD~54662 (and/or its possible binary companion) cannot be counted as an ionizing source for the nebula.

Remembering that the other O star HD~57682 is also far from the nebula (Table \ref{lymantable}), this means that in fact we are excluding the two runaway O stars from the nebula ionization budget.

As for the other two O stars, in projection (see Fig. \ref{bubble}) and by decreasing order of Lyman continuum flux, HD~53975 lies very close to the small, but bright, Sh~2-297 nebula and may be its main ionization source, but like HD~54662, on geometrical grounds it does not seem to contribute to the large scale  to the ionization of Sh~2-296. On the other hand, HD~54879 is definitely better located (it lies close to the centroid of the apparent arc associated with Sh~2-296), but it is the weakest of the four O stars, with a Lyman continuum flux at least 10 times smaller than HD~54662.

We conclude that, contrary to previous claims, the ionization of the Seagull nebula cannot be attributed entirely to the OB stars known in the area, but only to a certain fraction $\eta_\star$ of their total Lyman continuum flux. On the basis of the above discussion, we  consider that at most two O stars (HD~53975 and HD~54879) really contribute to this ionization, with a total of $N_{Lyc} = 4.8 \times 10^{48}$s$^{-1}$, i.e., $\sim 30\%$ of the measured flux $N_{Lyc}$. In other words, we estimate $\eta_\star \sim 0.3$ or less.

Consequently, another energy source is needed to sustain the remaining ($\gtrsim 70\%$) ionization of the nebula.

\subsection{X-ray heating as an ionization source?}

\begin{figure}
\centering
\includegraphics[width=0.43\textwidth,angle=0]{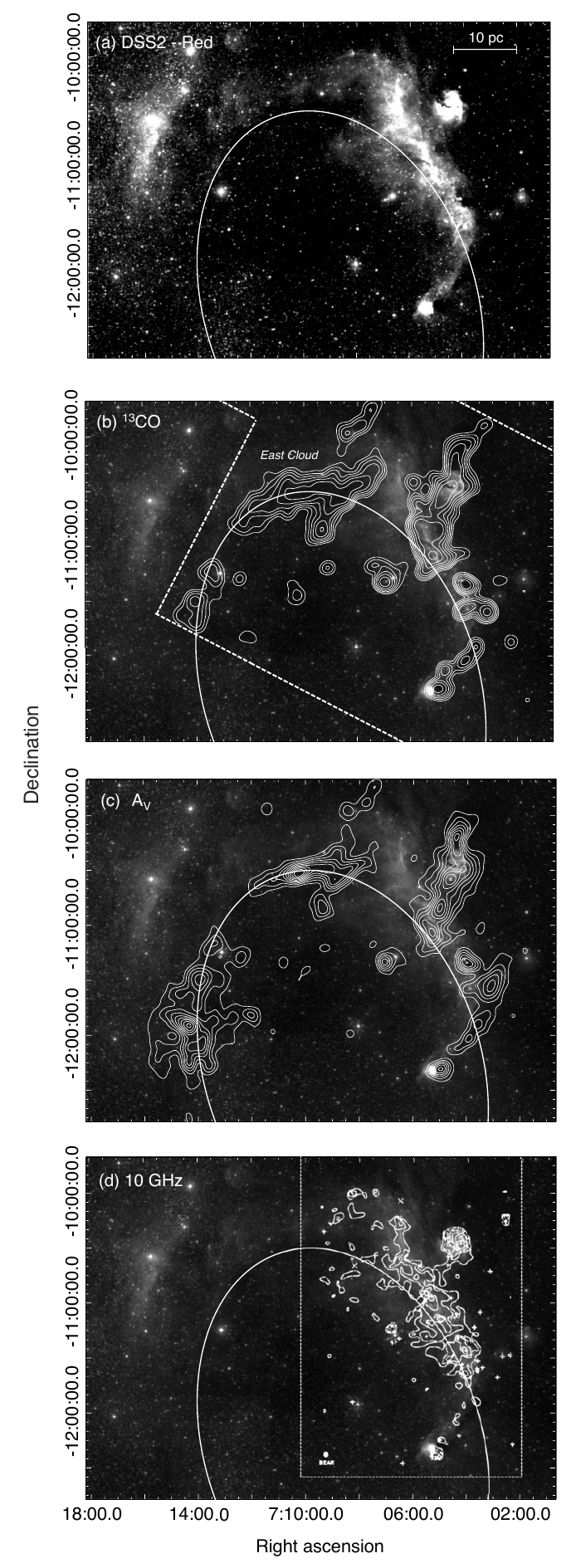}
\caption{(a) \emph{DSS2}-R image of the Sh~2-296 nebula (pixel size 1.7 arcsec). The ellipse outlines the proposed shape of the CMa R1 shell. The contours in each panel trace: (b) the $^{13}$CO emission (courtesy of the Osaka group, see \protect\citealt{Onishi2013}). The angular resolution (beam size) is 2.7 arcmin in HPBW; (c) the extinction ($A_V$) from the \emph{2MASS} maps of \protect\cite{Dobashi2011},  resolution of 1 arcmin; (d) the continuum radio emission at 10 GHz extracted from \protect\cite{Nakano1984}, beam size (HPBW) of 2.7 arcmin.}
\label{contours}
\end{figure}


As argued above, the CMa shell we see now is likely associated
with nested SNRs (like Barnard’s Loop), i.e., itself an SNR resulting from three successive explosions a few Myr apart.
Then, the prime consequence of the existence of a few-Myr old SNR, at that age slowly expanding in the ambient ISM in the course of a late radiative phase and visible in the form of a shell, is the \textit{X-ray emission} from a ``bubble'' of cooling plasma, previously heated by shock waves during the Sedov phase, and now bounded by this shell.

For the CMa shell, a possible scenario would then be as follows. 
The age $\tau$ of the shell would \textit{a priori} correspond to the time elapsed since the first explosion, i.e., the traceback time of the first runaway, HD 54662 ( $\tau = t_{runaway,1} \sim   6$ Myr). However, for the X-ray emission 
the \textit{cooling time}  $\tau_{cool}$ of the plasma would have to be at least comparable to the time elapsed since the last explosion, i.e., the traceback time of the third, fast runaway, HD 57682 ($\tau_{cool} \gtrsim t_{runaway,3} \sim  1 Myr $). Indeed, in spite of the fact that each explosion ``re-heats'' the previous SNRs, the youngest SNR is the one dominating the energetics (as shown by the recent theoretical
models by \cite{Krause2014a}; \cite{Krause2014b} and \cite{Yadav2017}; see discussion in Sect.6.4), thus essentially giving the CMa shell its current shape now. This is the
main reason why it is reasonable to ``stop'' the traceback trajectory of HD 57682 near the center of the shell, i.e., at $\sim 1$ Myr.

Such a scenario suggests an alternative (additional) energy source for the Sh 2-296 nebula, in terms of heating by X-rays from a $\sim 1$ Myr hot bubble
enclosed by the CMa shell, in contact with the nebula. This development is, however, beyond the scope of this paper and will be presented in a forthcoming
work where we will use the constrains provided by the stellar
runaway timescales and observed bubble parameters to
test this hypothesis and compare with X-ray observations.

\section{Discussion}
\label{sec:discussion}

\subsection{Star formation triggered by SN explosions?}
\label{starformation}

As mentioned above, the young stellar population of the CMa OB1 association contains few (late) O stars, dozens of B stars and hundreds low-mass young star candidates detected by X-ray observations \citep{Gregorio-Hetem2009,Santos-Silva2018}. Since we argue that the CMa shell is the result of at least three nested SNRs formed between $\sim 1$, 2 and $\sim 6$ Myr ago, it is worth discussing whether this young star population has any relationship with them.

Recently, \cite{Fischer2016} conducted a search for YSOs in a 100 deg$^2$ area around the CMa OB1 region, using data from the Wide-Field Infrared Survey Explorer ($WISE$). The authors were able to identify several groups of YSOs that are spatially distributed in a circular area, which approximately coincides with the  ring of optical and radio emission reported by \cite{Herbst10}, and which are in majority projected onto Sh~2-296. In Figure \ref{figysos}, we show that the distribution of these groups (see the yellow circles)  also coincides with our proposed elliptical shape of the CMa shell. Two groups, in particular, are projected onto the eastern border of the shell where no emission is seen in the images at any wavelength.


\begin{figure*}
\centering
\includegraphics[width=0.9\textwidth,angle=0]{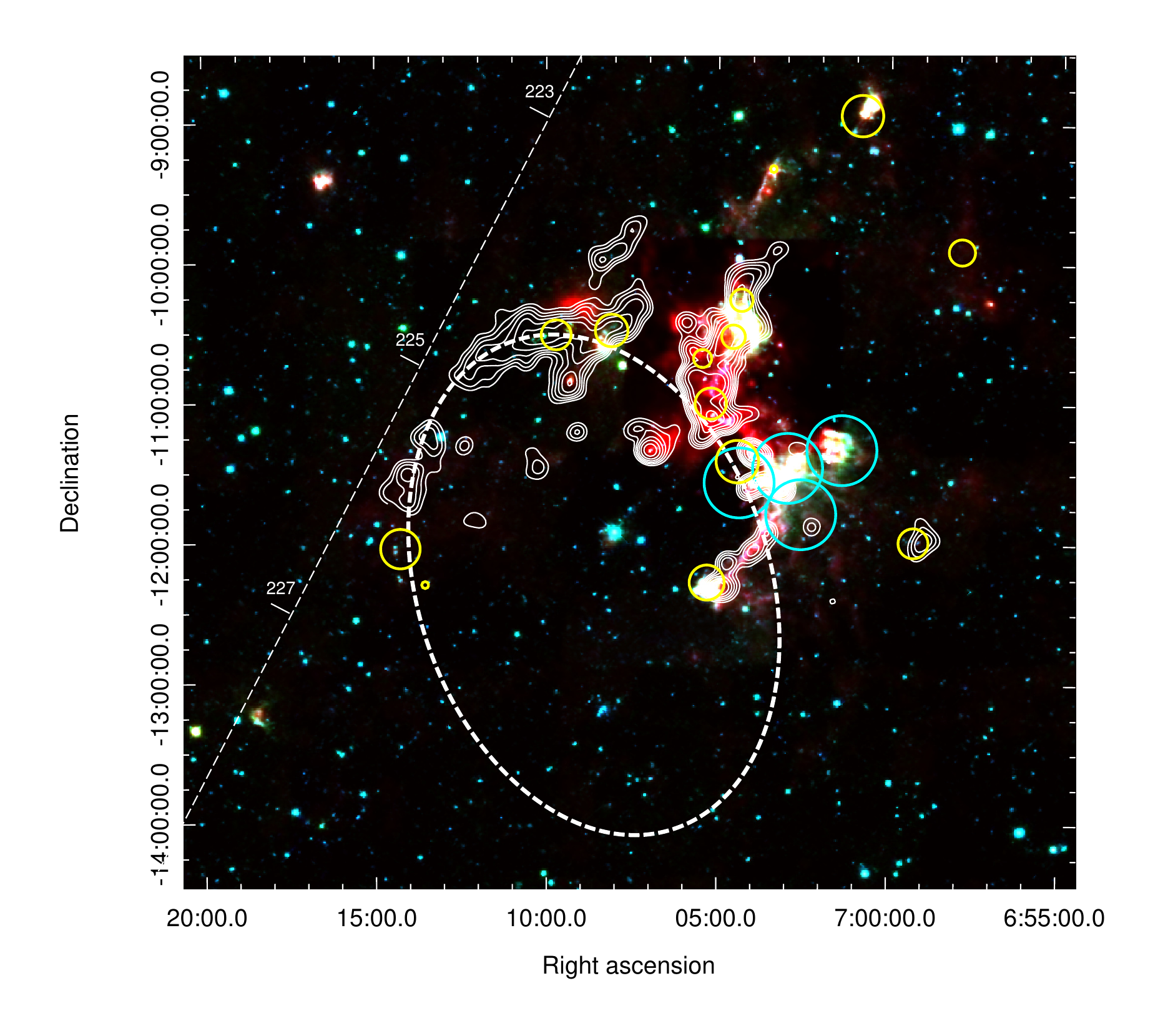}
\caption{\emph{WISE} composite color image of the CMa R1 region. {\it Red}: band 4, 22.2 $\mu$m. {\it Green}: band 3, 12.1 $\mu$m. {\it Blue}: band 1, 3.4 $\mu$m. The yellow circles delimit the area of the YSOs groups identified by Fischer et al. (2016). The circles in cyan show the position of {\it XMM-Newton} fields observed by  \cite{Santos-Silva2018}, 387 young stars are found in these fields. The majority have counterparts in the X-ray sources previously detected with \emph{ROSAT} by Gregorio-Hetem et al. (2009), though these indicate the presence of an extended distribution of young stars west of the nebula. The white contours trace the $^{13}$CO emission. The dashed thin line shows the galactic longitude at $b = 0^{\circ}$ and the white dashed ellipse represents the CMa shell (see Fig. \ref{bubble}). }
\label{figysos}
\end{figure*}


Even though the determination of ages in star forming regions is a complicated matter, several authors report an age spread of the young stellar population in CMa R1 \citep{Shevchenko1999, Gregorio-Hetem2009, Fischer2016,Santos-Silva2018}, containing groups of stars as old as $\sim 20$ Myr \citep[e.g.][]{Santos-Silva2018} and as young as $\sim 2$ Myr, as can be inferred by the predominance of Class I sources in some the groups identified by \cite{Fischer2016}.

Besides the shell structure and the apparent age mixing, other signatures of triggered star-formation are present in Sh~2-296, such as pillars and bright-rimmed clouds \citep{Rebull2013}, which are clearly visible in the H$\alpha$ image shown in Fig. \ref{halphazoom}. If the SN explosions that we inferred from the study of runaway stars occurred between $\sim$ 1 and  6 Myr ago, they could not be responsible for the formation of the oldest stars in the region, but could however have enhanced the most recent star-formation activity. For example, it is possible that the most recent SNR played a role locally in the formation of the youngest clusters of the association, such as the young stars in Sh~2-297 with an estimated age of 1 Myr \citep{Mallick2012}.

Therefore, CMa OB1 appears to be the relic of a spatially extended active star-forming region that has undergone the influence of its former massive star population (in particular at least three SN explosions), but, with only a few O stars remaining at the center of the shell, is now in the last stages of star formation activity. The CMa bubble itself could be considered as ``extinct'', since it is not powered any more by massive stars, but is quietly cooling by X-ray emission, while a fraction of this emission may go into heating and ionizing the Sh~2-296 nebula.

\subsection{Myr-old supernova remnants in the Galaxy: The CMa shell in context}
\label{context}

In the preceding sections, we have presented evidence that a large, $\sim 60$ pc diameter shell exists. It is, therefore, useful to put the CMa shell in context, i.e., comparing its properties with those of other large shell-like or bubble-like structures known in our Galaxy.

The closest case of a Myr SNR is undoubtedly the one surrounding the Sun. Indeed, there is evidence for a nearby ($d \sim$ 100-200 pc) SN explosion $\sim 2$ Myr ago, from excess short-lived radionuclides like $^{60}$Fe (half-life $\sim 2.6$ Myr) in the Earth's ocean crust \citep{Fry2015}, and from the anomalous $p/\bar{p}$ ratio in cosmic rays \citep{Kachelri2015}. This SNR is likely part of a local complex of interlaced ``bubbles'' \citep[the so-called ``Local Bubble'', e.g.,][]{Lallement2003}, and may be connected with the recently discovered, $\sim 2$ Myr old Antlia SNR \citep[][and refs. therein]{Tetzlaff2013}. We note that among the possible tracers of this SN explosion include the pulsar PSR J0630-2834 and, analogous to the CMa shell, a runaway (A-type) star: HIP~47155.

On a galactic scale, 61 HI ``supershells'' ($D > 100$ pc), first discovered by \cite{Heiles1984} are now known \citep{McClure2002}. As shown by these authors, some of them are found to lie at very large distances from the Sun (up to $\sim 20$ kpc), and may have extremely large sizes ($D \sim 700$ pc), reached in a low-density interstellar medium after several 10 Myr, with expansion energies $\sim 10^{51}-10^{53}$ erg.\footnote{However, as shown recently by Suad et al. (2014, 2019)
a much larger population of $\sim 500$ galactic supershells exists, but with
smaller to much smaller kinetic energies ($10^{47}$ to $\sim 10^{51}$ erg).}

Their exact origin is not clear, likely a combination of many SN explosions and galactic-arm structure. One of them, GSH 138-01-94, has been found to be ``the largest and oldest SNR'', with a diameter of $D = 360$ pc (at a distance $d = 16.6$ kpc), and an estimated age $\sim 4.3$ Myr \citep{Stil2001}. By contrast, smaller supershells of the same age do exist closer to the Sun, like GSH 90-28-17 \citep[$D = 132$ pc, $d = 400$ pc, age $\sim 4.5$ Myr:][]{Xiao2014}. 

Very large, filled structures known as ``superbubbles'' are also known, like the famous X-ray emitting Orion-Eridanus bubble ($D = 320$ pc for a distance $d = 400$ pc), which is likely a complex of nested SNRs including Barnard's Loop, with an estimated age $\simeq 8$ Myr \citep{Ochsendorf2015, Pon2016}. 

So we must add another criterion for comparison with the CMa shell: to be bordered by molecular clouds (at least in part).
 A first example is the Carina region, in which two colliding supershells have accumulated interstellar material so as to form a giant molecular cloud in between \citep{Dawson2015}. One of the supershells, GSH~287+04-17, is seen in HI; it is located $2.6 \pm 0.4 $ kpc from the Sun, and had a size $150 \times 230$ pc, so reaches high galactic latitudes. The other one is associated with the Carina OB2 association; its distance is $2.9 \pm 0.9$ kpc (so is compatible with that of the HI supershell), and has a size $80 \times 30$ pc. Both supershells have estimated input energies $E \sim 5 \times 10^{51}$ ergs, and ages less than $\approx$ a few Myr. To compare with the CMa situation, we point out that \cite{Gregorio-Hetem2009} analyzed the spatial spread of the stellar population associated with the CMa molecular clouds, finding that young stars (ages $< 5$ Myr) were present on the East and West sides of the clouds (see above, Sect.1). This result led them to suggest that these stars might had been formed as a result of SN explosions ``squeezing'' the molecular clouds. However, the inspection of \emph{SHASSA} or \emph{DSS} R maps do not reveal evidence for a shell on the West side of the CMa molecular clouds, so the CMa shell (to the East) appears ``single'' compared to the Carina pair of supershells.

Another case is the G349.7+0.2 SNR, discovered in radio continuum with the VLA thanks to its non-thermal spectrum. According to \cite{Reynoso2001}, this SNR is in contact with several molecular clouds, as testified by the detection of 1617 MHz OH masers; furthermore, it seems to have triggered star formation, in the form of an ultra-compact HII region. It has a diameter $D \simeq 100$ pc, and an age $\sim 4$ Myr, so it appears more similar to the CMa shell. But it lies very far,  at $d = 23$ kpc from the Sun.

Altogether, the CMa shell appears, at least observationally, as somewhat unique among large, Myr galactic SNR in contact with molecular clouds.{\footnote{Many other cases of interactions between SNRs and molecular clouds are known in the Galaxy, but they concern much younger ($\approx 10^4-10^5$ yrs old) SNRs, in particular linked with GeV-TeV $\gamma$-ray sources \citep[e.g.,][and refs. therein]{Gabici2015,Vaupre2014}.} It is quite close to the Sun ($d \sim 1$ kpc) and in addition its association with the local stellar population is well studied  (see Sect. \ref{starformation}).

\subsection{A CMa superbubble ?}\label{sec:superbubble}

On the large scale H$\alpha$ image shown in Appendix \ref{sec:gallery} we note a diffuse, faint ``supershell'' of emission (larger dotted elliptical ring), which has no counterpart at other wavelengths, except perhaps in the radio range (21cm and 408 MHz). These features are brought up in Fig. \ref{arcs}. We can clearly see that, beyond the shell we have discussed up to now (and which is then overexposed), there exists a much larger, diffuse structure of H$\alpha$ emission, i.e., a supershell $\sim 140$ pc in size. These features are also  apparent in the {\it DSS} image shown in Fig. \ref{bubble}.

We have an approximate elliptical fit to the supershell not only on Fig. \ref{arcs}, but also on our multi-wavelength gallery (Fig. \ref{gallery}): contrary to the CMa shell, this supershell is not seen, so the properties of the ISM in which it must be embedded cannot be known.}

We are led to suggest that these features must be the dim remnants of SN explosions having taken place {\it before} the three which we have associated with the CMa shell, and that the H$\alpha$ supershell might similarly enclose a hot superbubble. It is interesting to note that the West border of this supershell coincides with the location of the older (10 Myr) group of X-ray sources detected by {\it ROSAT} around GU CMa \citep{Gregorio-Hetem2009}.

As discussed in the preceding section, other structures, comparable in size to the CMa shell (though usually larger, with D $\sim$ 100 pc) and referred to as ``superbubbles'', have been detected in X-rays long ago and thus may be targets to look for their heating effects on the surrounding gas. However, these superbubbles are associated with clusters of O stars, and are ``pressure-driven'', i.e., are now expanding under the influence of stellar winds and/or recent SN explosions, contrary to the CMa bubble. Said differently, their X-ray pressure is found to exceed the HII gas pressure by a significant factor, and/or they are characterized by large expansion velocities ($v_{exp} \sim$ several 10 km~s$^{-1}$).

Perhaps the most famous known concentration of superbubbles lies in the LMC, as shown by observations dating back from the late seventies, although at the time they were simply called SNRs \citep[e.g., ][]{Lasker1977}. Early {\emph{Einstein}} observations revealed the presence of many previously unknown X-ray emitting superbubbles, of diameter $\sim 100$ pc or more, and X-ray luminosities $L_X \sim 1-10 \times 10^{35}~{\rm erg~s}^{-1}$ \citep[e.g., ][]{Chu1990}, so in the same range as the predicted X-ray luminosity of the CMa shell alone.

Deeper observations of the LMC were obtained with \emph{ROSAT}, revealing new superbubbles \citep[e.g., ][]{Dunne2001}. Their sizes are $D \sim 100-200$ pc, and their expansion velocities $v_{exp} \sim 20-70~{\rm km~s}^{-1}$. The X-ray spectral fittings reveal soft X-ray emission ($kT \sim 0.2-0.9$ keV, or $T \sim 2-9$ MK; $n_H \sim 0.01-0.1~{\rm cm}^{-3}$), confirming X-ray luminosities $L_X \sim 10^{35-36} ~{\rm erg~s}^{-1}$. Based on their sample of 13 superbubbles, \cite{Dunne2001} found a correlation with the richness and age of the exciting OB associations, thus interpret the origin of X-rays as coming from stellar winds, enhanced by SNRs, resulting in ``pressure-driven'' superbubbles. The link with the young, massive stellar population suggested an age on the order of a few Myr.

More detailed observations were obtained by {\emph{XMM-Newton}} on the N158 \citep{Sasaki2011} and N206 \citep{Kavanagh2012} HII regions.  N156 appears as a somewhat extreme superbubble, with a harder spectrum ($kT \sim 1$ keV) and high pressure ($P/k \sim 10^6 ~{\rm cm}^{-3}$ K), and a corresponding high X-ray luminosity ($L_X = 2.3 \times 10^{38}~{\rm erg~s}^{-1}$), i.e., two orders of magnitude larger than for typical LMC superbubbles. This region is similar to the 30 Dor star forming region, with a large population of very massive stars. The age estimate is very young ($\sim 1$ Myr), but the high $L_X$ implies that 2-3 supernovae may have already exploded. By comparison, N206 appears more typical but still hot ($kT \sim 1$ keV), with $P/k \simeq 5 \times 10^5~{\rm cm}^{-3}$ K. The energy content is $E \sim 3 \times 10^{51}$ erg. The stellar population is poorly known, but again massive stars are likely to explain the formation of the superbubble, with a correspondingly young age.


\begin{figure}
\centering
\includegraphics[width=0.48\textwidth,angle=0]{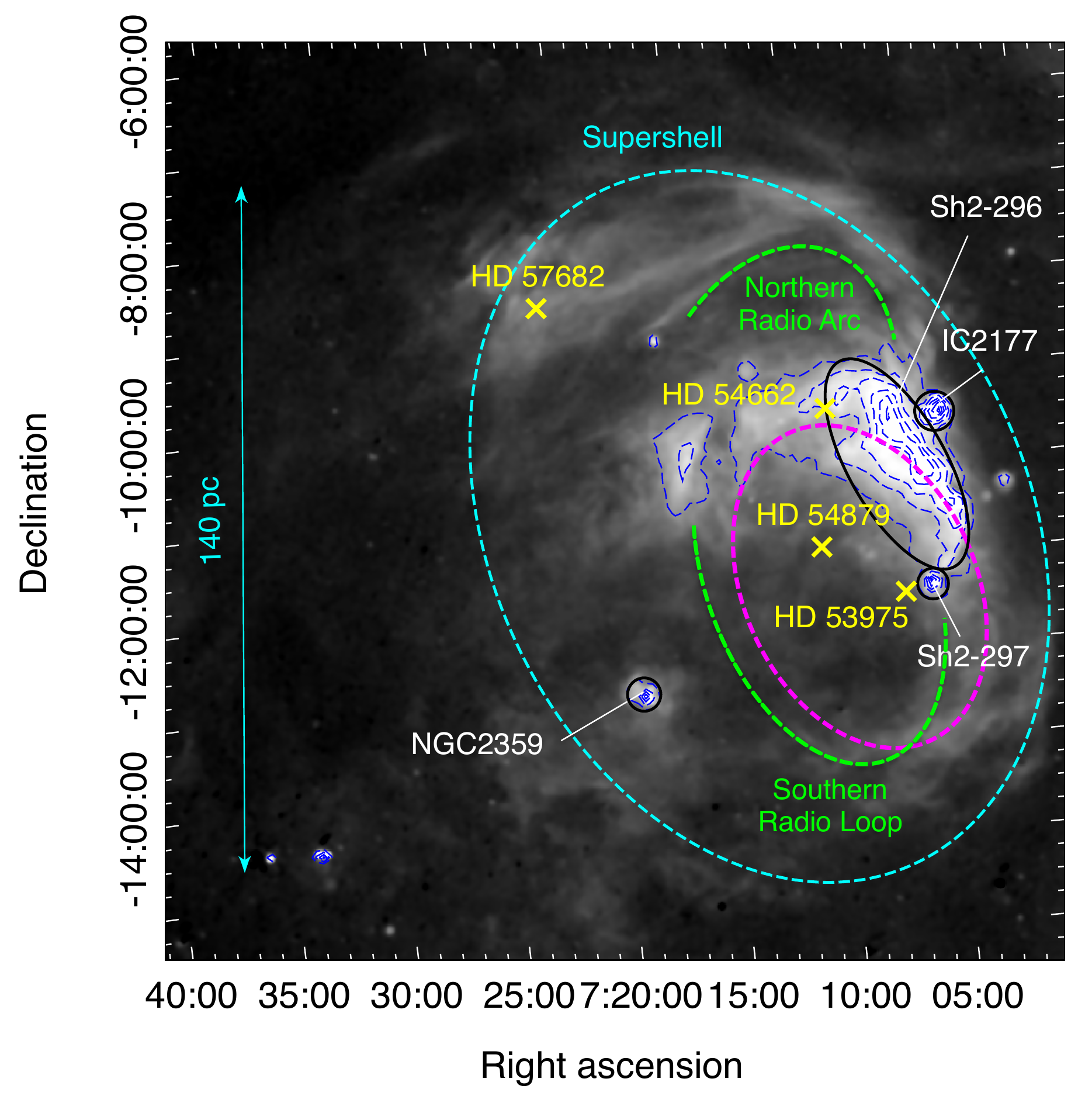}
\vspace{-0.2cm}
\caption{The large-scale H$\alpha$ environment of the CMa shell, from the \emph{SHASSA} survey, enhancing the faint emission, down to a surface brightness of $\sim 150$ dR. Conspicuous H$\alpha$ filaments can be seen, well correlated with the radio map of Gaylard \& Kemball (1984) (their Fig. 1), in particular with features that we outline in green on this Figure, named "Northern Arc" and "Southern Loop" (this one well correlated with the southern half of the CMa shell). We suggest that this faint H$\alpha$ emission is associated with a ``superbubble'' $\sim 140$ pc in diameter (here outlined in the form of a dashed cyan ellipse), analog to Barnard's Loop around the Orion nebula.} 
\label{arcs}
\end{figure}


Of relevance to our work, the conclusion emerging from these studies is that, as a rule, superbubbles in the LMC expand and are sustained by winds and SN explosions of massive stars currently present in OB associations. As we have argued, this is not the case of the CMa shell, which is not expanding, and contains only a few late-type O stars, having weak winds and insufficient UV radiation to ionize the Sh~2-296 nebula. So in that sense, the CMa case may be also unique among superbubbles.

\section{Summary and Concluding Remarks}
\label{conclusions}

We have found strong evidence in favor of connections between the Sh~2-296 nebula and a complex of nested $\sim$ 1 - 6 Myr-old SN remnants. First, a large ($D \sim 60$ pc) shell
(the ``CMa shell'') is visible at various wavelengths (optical, HI, CO, etc.)  but has apparently escaped detection until now because of its large angular size (over 80 sq. deg.). Second, three runaway stars are found in association with the nebula: using proper motions measured by \emph{Gaia} and {\it Hipparcos} and radial velocities, we traced back the position of these stars, and we find that they have been likely ejected from approximately the same location, inside the shell in (at least) three successive SN explosions $\sim$ 6 Myr, 2 Myr and 1 Myr ago.

On the other hand, after having discussed the Lyman continuum flux and the spatial distribution of the O stars in the region, we conclude that they cannot fully account for the ionization the nebula derived from radio and H$\alpha$ observations. 

In accordance to our multiple SN explosion scenario, we suggest that the Sh~2-296 nebula may be heated by X-rays from a hot bubble bound by the CMa-shell.

The region may actually have had a more complex history, since we also find evidence that the CMa bubble is itself nested in a larger, ionized structure, visible in H$\alpha$ surveys and bounding a ``superbubble'' $\sim 140$ pc in size.
Indeed, while the three SN successive explosions advocated in this paper may have played a role in inducing recent star formation in CMa at least locally, the older population of low mass stars (> 10 Myr), found by \cite{Gregorio-Hetem2009} and \cite{Santos-Silva2018} cannot be explained by them. However, the membership and age of a significant part of these objects have been confirmed by the parallaxes from \emph{Gaia DR2}, as discussed in Appendix C, leading us to suggest that they may be causally related to the existence of the H$\alpha$ supershell as a relic of previous SN explosions. In other words, at least in a global sense, several localized episodes of star formation in the CMa~OB1 association would ``echo'' the successive supernova explosions having occurred in the region in the past over several Myr.

Then, the fact that there are only a few massive stars left (with a lifetime of several Myr only), implies that the CMa OB1 association that we are witnessing now is the ``swan's song'' of the history of a formerly very active, $\sim 10$ Myr old star-forming region.

\section*{Acknowledgements}

\addcontentsline{toc}{section}{Acknowledgements}

BF acknowledges financial support from CNPq (projects: 205243/2014-2/PDE and 150281/2017-0). JGH thanks CAPES/Cofecub (712/2011), CNPq (305590/2014-6), and FAPESP (2014/18100-4).  TSS thanks CAPES (Proj: PNPD20132533). We thank T. Onishi (Osaka University) for having provided us with his CO data in advance of publication. TM thanks S. Guisard for generously providing the wide-field H$\alpha$ picture of the Seagull Nebula used in this paper. The authors would also like to thank S. Federman, P. Boiss\'e, J. R. D. L\'epine and P. A. B. Galli for many productive discussions and feedback on this manuscript. This work has made use of the VizieR, SIMBAD and Aladin databases operated at CDS, Strasbourg, France. This work has made use of data from the European Space Agency (ESA) mission
{\it Gaia} (\url{https://www.cosmos.esa.int/gaia}), processed by the {\it Gaia}
Data Processing and Analysis Consortium (DPAC,
\url{https://www.cosmos.esa.int/web/gaia/dpac/consortium}). Funding for the DPAC
has been provided by national institutions, in particular the institutions
participating in the {\it Gaia} Multilateral Agreement.

\bibliographystyle{aa}
\bibliography{cmashell}


\begin{appendix}

\section{Multi-wavelength view of the CMa Shell}\label{sec:gallery}

In this Appendix, we present a gallery of the CMa R1 region in different wavelengths.  For comparison with Fig. \ref{arcs},  the position of the four O type stars associated with the region is marked by crosses. These images demonstrate, as mentioned in Sect. \ref{sec:runaways}, that Sh~2-296 is part of a large elliptical ring that extends beyond the familiar bright arc-shaped nebula seen in the optical images. This ring, or shell-like structure, is specially clear in the H$\alpha$ image of the region that we present here again (see also Fig. \ref{arcs}) for comparison with other wavelengths. 
These images (Fig. \ref{gallery}) can help us better constrain the shape of the shell, tracing the eastern and southern parts of the structure. It is interesting to note that no emission at these wavelengths is detected within the region surrounded by the shell, which also seems to be mostly devoid of young stars (see Fig. \ref{figysos}). Even considering the changes in resolution, it can be also noted that  the large structure of the shell remains roughly  traced. Comparing, for instance, the lower resolution found in the images from Planck LFI and {\it Haslam} 408~MHz with the better defined H$\alpha$ image, we see the same overall distribution traced by the emission at these different wavelengths.


\begin{figure*}
\centering
\includegraphics[width=0.8\textwidth,angle=0]{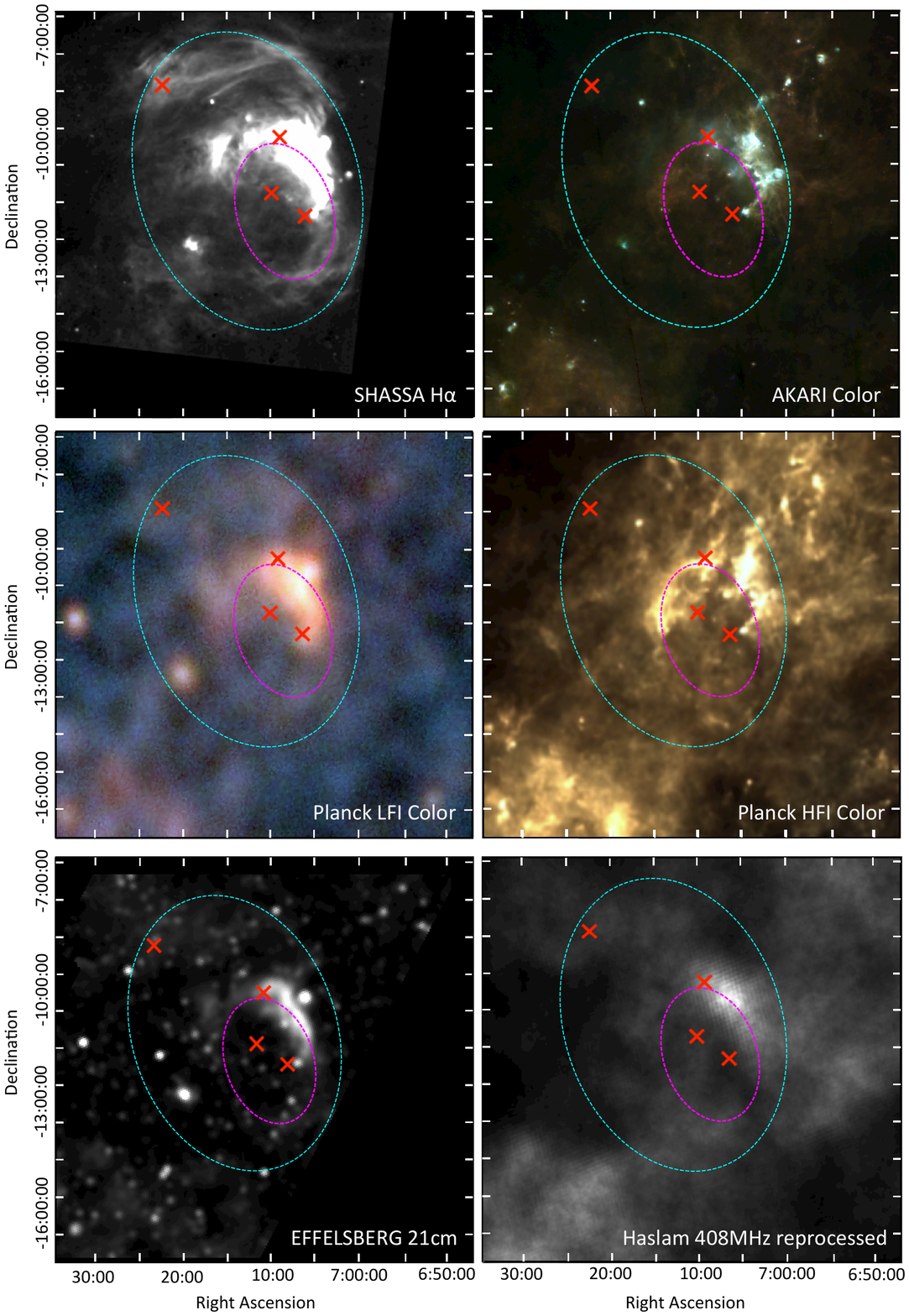}
\caption{Images of CMa R1 at different wavelengths obtained with Aladin Progressive Sky (HiPs). The position of the O stars HD 57682, HD 54662, HD 54879 and HD 53974 is marked as red crosses. The small magenta dotted ellipse represents the ``CMa shell'', and the large cyan dotted ellipse is enclosing the superbubble suggested by us. From top left to bottom right:  H$\alpha$ image from the \emph{SHASSA} Survey (see also Fig. \ref{arcs}); \emph{Akari} \citep{Murakami2007} color-composition at 140 $\mu m$ (red), 90 $\mu m$ (green) and 65 $\mu m$ (blue); \emph{Planck} Release 2 \citep{Planck2016} HFI color-composition at 353 GHz (red), 545 GHz (green), 857 GHz (blue);  \emph{Planck} Release 2 LFI color-composition at 30 GHz (red), 44 GHz (green) and 70 GHz (blue);  image from the {\it Effelsberg} telescope  at 21 cm \citep{Reich1990}; and from the reprocessed {\it Haslam} survey at 408 MHz \citep{Remazeilles2015}. Note the correspondence between the \emph{Planck} LFI emission (cold dust) and the $^{13}$CO emission shown on Fig.\ref{bubble}.
}
\label{gallery}
\end{figure*}


\section{Proper motion of runaway stars} \label{sec:propermotion}

To derive the peculiar, or local, proper motion of the stars, i.e. the proper motion of the star corrected for solar motion and galactic rotation, it is convenient to first convert the observed proper motion components from the equatorial (${\mu}_\alpha \cos \delta \textrm{ and } \mu_{\delta}$) to the galactic (${\mu}_l \cos b \textrm{ and } \mu_b$) coordinate system. To achieve this, we follow the transformation derived by \cite{Poleski2013}, see also \cite{Johnson1987}:

\begin{equation}
\left(\begin{array}{c}
{\mu}_l \cos b \\
\mu_b \end{array} \right)= \frac{1}{\cos b} 
\left(\begin{array}{cc}
C_1 & C_2  \\
-C_2 & C_1 \end{array} \right)
\left(\begin{array}{cc}
{\mu}_\alpha \cos \delta \\
\mu_{\delta} \end{array} \right)
\end{equation}

\noindent where ($l,b$) and ($\alpha, \delta$) are the galactic and equatorial coordinates of the star, respectively. The coefficients $C_1$ and $C_2$ are given by:

\begin{equation}
C_1 = \sin {\delta_G} \cos {\delta} - \cos {\delta_G} \sin {\delta} \cos{(\alpha - {\alpha}_G)}
\end{equation}

\begin{equation}
C_2 = \cos {\delta_G} \sin{(\alpha - {\alpha}_G)}
\end{equation}

\noindent and ${\alpha}_G = 192^{\circ}.85948$ and ${\delta}_G = 27^{\circ}.12825$ are the equatorial coordinates of the North galactic pole.

In this Appendix, we will find the resulting proper motion and velocities of the runaway star HD~53974, as an example. Using Eq.(B1) with the coefficients given by Eqs. (B2) and (B3), we have that the corresponding values of the proper motion of HD~53974, expressed in galactic coordinates, are ${\mu}_l \cos b = (-4.37 \pm 0.4)$ mas yr$^{-1}$ and $\mu_b = (-1.26 \pm 0.9)$ mas yr$^{-1}$.

The proper motion components of the peculiar motion of the star can then be written as the observed proper motion minus the components due to the basic solar motion  (${(\mu_l \cos b)}_{\odot} , {(\mu_b)}_{\odot}$) and galactic rotation (${(\mu_l \cos b)}_{rot} , {(\mu_b)}_{rot}$), as described in  \cite[see][]{scheffler1988, moffat1998}:

\begin{equation}
{(\mu_l)}_{pec} \cos b=  (\mu_l \cos b) - {(\mu_l \cos b)}_{\odot} - {(\mu_l \cos b)}_{rot}
\end{equation}

\begin{equation}
{(\mu_b)}_{pec} =  (\mu_b) - {(\mu_b)}_{\odot} - {(\mu_b)}_{rot}
\end{equation}

\noindent where the terms due to the Basic solar motion can be written as 

\begin{equation}
 {(\mu_l \cos b)}_{\odot} = \frac{1}{Kr}\left(U_{\odot} \sin l - V_{\odot} \cos l \right)
\end{equation}

\begin{equation}
{(\mu_b)}_{\odot} =  \frac{1}{Kr}\left(U_{\odot} \cos l sin b + V_{\odot} \sin l \sin b - W_{\odot} \cos b\right)
\end{equation}

\noindent  where $U_{\odot}, V_{\odot}, W_{\odot} $, are the components of the solar peculiar motion in the direction toward the galactic center, the direction of the galactic rotation, and the north galactic pole, respectively. For proper motions given in [mas yr$^{-1}$], $K = 4.74047$ km  yr s$^{-1}$ is the ratio between the astronomical unit in km and the number of seconds in a year. 

We adopt a model where the stars move on the galactic plane in circular orbits with radius $R$ around the galactic centre, in a flat rotation curve. If $R_o$ is the Solar galactocentric distance and $r$ is the distance from the star to the Sun, then the galactocentric radius $R$ of the star can be expressed in terms of $r$ by means of the cosine rule \citep[see][]{scheffler1988}

\begin{equation}
R^2 =R_o^2 +r^2 {\cos}^2{b} - 2r{R_o}\cos{b} \cos{l}
\end{equation}

The angular velocity of the orbital motion will then be denoted by

\begin{equation}
\omega (R) = \frac{V}{R} \textrm{ and } {\omega}_o = \frac{V_o}{R_o}
\end{equation}

The proper motion terms due to the galactic rotation are given as

\begin{equation}
{(\mu_b)}_{rot} = -\frac{1}{K}\left[\frac{R_o}{r} (\omega - \omega_{o}) \sin l \sin b \right]
\end{equation}

\begin{equation}
{(\mu_l)}_{rot} = \frac{1}{K} \left[\frac{R_o}{r \cos b} (\omega - \omega_{o}) \cos l - \omega \right]
\end{equation}

Assuming that HD~53974 is at the same distance as CMa OB1 ($r = 1$ kpc), we find $R=8.74$. Using this value in Eq.(B9), we have $\omega = 26.32$. We adopt (U$_{\odot}$, V$_{\odot}$ , W$_{\odot}$) = (11.1, 12.24, 7.25) km s$^{-1}$ \citep{Schonrich2010} and a flat rotation curve with R$_0$ = 8 kpc and V$_0$ = 230 km s$^{-1}$. From Eqs. (B10) and (B11), we obtain ${(\mu_l \cos b)}_{rot}$ = -2.63 mas yr$^{-1}$  and ${(\mu_b)}_{rot}$ = 0.09 mas yr$^{-1}$, and from Eqs. (B6) and (B7) we obtain ${(\mu_l \cos b)}_{\odot}$ = 0.19  mas yr$^{-1}$ and ${(\mu_b)}_{\odot}$ = -1.43 mas yr$^{-1}$. 

The resulting peculiar proper motion components are then

\begin{equation}
{(\mu_l)}_{pec} \cos b= (\mu_l \cos b) - {(\mu_l \cos b)}_{\odot} - {(\mu_l \cos b)}_{rot} 
\end{equation}

${(\mu_l)}_{pec} \cos b$ = (-1.94 $\pm$ 0.16) $\textrm{ mas yr}^{-1}$

\begin{equation}
{(\mu_b)}_{pec} =  (\mu_b) - {(\mu_b)}_{\odot} - {(\mu_b)}_{rot} = (0.06 \pm 0.68) \textrm{ mas yr}^{-1}
\end{equation}

The position angle $\theta$ of the peculiar proper motion with respect to the North galactic pole, counted as positive in the direction of increasing galactic latitude, is

\begin{equation}
\theta = {\tan}^{-1} \left(\frac{{(\mu_l)}_{pec} \cos b}{{(\mu_b)}_{pec}}\right) = 1.87^{\circ} \pm 29.7^{\circ}
\end{equation}

Correcting the heliocentric radial velocity (RV) for basic solar motion and galactic rotation, the components of the peculiar radial velocity of the star can be written as

\begin{equation}
({v_r})_{pec} =  RV - ({v_r})_{\odot} - ({v_r})_{rot}
\end{equation}

\noindent where

\begin{equation}
({v_r})_{\odot} = -U_{\odot} \cos l cos b - V_{\odot} \sin l \cos b - W_{\odot} \sin b = 16.72 \textrm{ km s}^{-1}
\end{equation}

\noindent and

\begin{equation}
({v_r})_{rot} = {R_o} (\omega - \omega_{o}) \cos b \sin l = 13.67 \textrm{ km s}^{-1}
\end{equation}

\noindent The velocity on the plane of the sky is given by the corrected peculiar proper motion:

\begin{equation}
({v_t})_{pec} = Kr \sqrt{({(\mu_l}_{pec} \cos b)^2 + {(\mu_b)}_{pec}^2} = 9.21 \textrm{ km s}^{-1}
\end{equation}

\noindent and the total space velocity of the star in relation to the local interstellar medium is then:

\begin{equation}
(v_{\star}) = \sqrt{({v_r})_{pec}^2 + ({v_t})_{pec}^2} = 12.28 \pm 4.84 \textrm{ km s}^{-1}
\end{equation}

In Table \ref{pmres}, we give the peculiar proper motions estimated by us for the three runaway stars found in CMa OB1. Comparison with the literature was only possible for HD~54662 (HIP ~34536) that was studied by \cite{Tetzlaff2011}  who found total peculiar velocity $v_{\star}$ = 33.7 km s$^{-1}$,  in good agreement with our result $v_{\star}$ = 36.3$\pm$ 2.6 km s$^{-1}$. Our other two runaways,
HD 53974 and HD 57682, are not present in the catalogue of \cite{Tetzlaff2011}.

\section{Ages of the CMa stellar populations after Gaia DR2} \label{sec:gaiaages}

The CMa young star candidates that were revealed by analyzing the near-IR counterparts of X-ray sources associated with the Sh2-296 nebula \citep{Gregorio-Hetem2009, Santos-Silva2018} can now have a refined estimate of their age and mass. The membership confirmation of these sources is an ongoing study that our team is currently performing based on parallaxes provided by \emph{Gaia DR2}, which will be discussed in a forthcoming paper.
As mentioned in the concluding remarks (Sect. \ref{conclusions}), the preliminary results of particular interest for the present work are those for sources showing more than 10 Myr in the catalogue from \cite{Santos-Silva2018}. Several are background or foreground objects. But almost half of them can be confirmed as CMa members. It is interesting to note that there is a trend of these objects (25/31) at distances in the range 1100-1250 pc that means an underestimate of absolute magnitude ($\sim 0.2 - 0.5$ mag). This difference would give a younger age for these sources, but not too different of the 10 Myr range. A more detailed analysis (considering individual values of $A_V$, for instance) is required to improve the age estimate for all the confirmed members associated with this star-forming region. Nevertheless these preliminary results confirm the concluding remarks concerning the presence of an older ($\sim 10$ Myr) population in CMa, possibly triggered by early SN explosions generating the H$\alpha$ supershell we have found.

\end{appendix}

\end{document}